\documentclass[fleqn,usenatbib]{mnras}

\usepackage{newtxtext,newtxmath}
\usepackage[T1]{fontenc}
\DeclareRobustCommand{\VAN}[3]{#2}
\let\VANthebibliography\thebibliography
\def\thebibliography{\DeclareRobustCommand{\VAN}[3]{##3}\VANthebibliography}

\usepackage{amsmath}
\usepackage{graphicx}
\usepackage{enumitem}
\usepackage{xcolor}
\usepackage{hyperref}
\usepackage{gensymb}
\usepackage{CJK}
\usepackage{tabularx}
\usepackage{orcidlink}
\usepackage{fontawesome}
\usepackage[normalem]{ulem}
\hypersetup{
    unicode, 
    colorlinks=true,
    linkcolor=linkcolor,
    citecolor=linkcolor,
    filecolor=linkcolor,
    urlcolor=linkcolor,
}
\definecolor{linkcolor}{rgb}{0.0,0.3,0.5}

\newcommand{\changed}[1]{{{#1}}}

\newcommand{\changes}[1]{#1}

\title[A golden thread through time and space]{The chemodynamical memory of a major merger in a NIHAO-UHD Milky Way analogue -- I. A golden thread through time and space}

\author[S. Buder et al.]{Sven Buder,$^{1}$\thanks{E-mail: sven.buder@anu.edu.au}\thanks{Australian Research Council DECRA Fellow}\orcidlink{0000-0002-4031-8553}
Tobias Buck,$^{2,3}$\orcidlink{0000-0003-2027-399X}
Ása Skúladóttir,$^{4}$\orcidlink{0000-0001-9155-9018}
Melissa Ness,$^{1}$\orcidlink{0000-0001-5082-6693}
Madeleine McKenzie,$^{5}$\thanks{NASA Hubble Fellow}\orcidlink{0000-0002-1715-1257}
and\newauthor
Stephanie Monty$^{6, 7, 8}$\orcidlink{0000-0002-9225-5822}
\\
$^{1}$Research School of Astronomy and Astrophysics, Australian National University, Canberra, ACT 2611, Australia\\
$^{2}$Universit{\"a}t Heidelberg, Interdisziplin{\"a}res Zentrum f{\"u}r Wissenschaftliches Rechnen, Im Neuenheimer Feld 205, D-69120 Heidelberg, Germany\\
$^{3}$Universit{\"a}t Heidelberg, Zentrum f{\"u}r Astronomie, Institut f{\"u}r Theoretische Astrophysik, Albert-Ueberle-Straße 2, D-69120 Heidelberg, Germany\\
$^{4}$Dipartimento di Fisica e Astronomia, Universitá degli Studi di Firenze, Via G. Sansone 1, I-50019 Sesto Fiorentino, Italy\\
$^{5}$The Observatories of the Carnegie Institution for Science, 813 Santa Barbara Street, Pasadena, 91101, CA, USA\\
$^{6}$Center for Interdisciplinary Exploration and Research in Astrophysics (CIERA), Northwestern University, 1800 Sherman Avenue,
Evanston, IL 60201, USA\\
$^{7}$Department of Astronomy, New Mexico State University, Las Cruces, NM 88003, USA\\
$^{8}$Institute of Astronomy, University of Cambridge, Madingley Road, Cambridge CB3 0HA, UK}

\date{Accepted 2026 June 5. Received 2026 May 31; in original form 2025 October 12}

\pubyear{\the\year{}}

\begin{document}
\label{firstpage}
\pagerange{\pageref{firstpage}--\pageref{lastpage}}
\maketitle

\begin{abstract} 
Understanding how past major mergers shaped the Milky Way's present-day structure is a key goal of Galactic archaeology. The Galaxy's chemical and dynamical structure retains the imprint of such events, including a major accretion episode around 8--10 Gyr ago. Recent findings suggest that present-day orbital energy correlates with stellar chemistry and birth location within the merging progenitor galaxy. Using a high-resolution NIHAO-UHD cosmological zoom-in simulation of a Milky Way analogue, we trace the birth positions, ages, and present-day orbits of stars accreted in its last major merger. We show that stars born in the progenitor's core are more tightly bound to the Milky Way and more chemically enriched, while those from the outskirts are less bound and more metal-poor. This supports the Sk\'ulad\'ottir et al.\ (2025) scenario \changes{that accreted progenitor stars of different chemistry were deposited onto different orbital energies as the galaxy was stripped from the outside in}, now in a cosmological context. Quantitatively, we measure a metallicity gradient with progenitor birth radius of $\mathrm{d[Fe/H]}/\mathrm{d}R_\mathrm{birth}^\prime \approx -0.05\,\mathrm{dex\,kpc^{-1}}$, demonstrating that abundance patterns retain measurable memory of formation location within the disrupted satellite. This chemodynamical memory is also evident in elemental planes such as [Al/Fe] vs. [Mg/Mn], consistent with gradients in progenitor star formation efficiency. We further show that common integrals-of-motion selections systematically miss stars from the chemically enriched core, biasing reconstructions toward the metal-poor outskirts. Together, our results demonstrate that chemodynamical memory survives the merger and can reconstruct the accreted galaxy's internal structure, while highlighting biases in current selections of accreted stars.
\end{abstract}



\section{Introduction}
\label{sec:introduction}

Understanding what shaped the Milky Way and its present-day structure remains a central goal of Galactic archaeology \citep{FreemanBlandHawthorn2002,BlandHawthorn_Gerhard2016}. The discovery of the thin and thick stellar disc components \citep{Yoshii1982, Gilmore1983}, characterised by their largely distinct spatial, chemical and age distributions \citep[for example][]{Bensby2014, Hayden2015}, has led to numerous formation scenarios being proposed. These include turbulent high-redshift star formation producing a kinematically hot thick disc \citep{Bird2013, Stinson2013}, internal heating and radial migration within an evolving disc \citep{Schoenrich2009, Minchev2013, Sharma2021b}, a strong sensitivity of star formation to temporarily lower gas accretion \citep{Grand2018, Orkney2026b}, as well as gas-rich mergers \citep{Brook2004, Buck2020} and the build-up of the disc under a two-infall scenario \citep{Chiappini1997, Spitoni2019}. More and more likely it seems to actually be a combination of multiple pathways \citep[e.g.][]{Orkney2026b}.

Thanks to new observations of the \textit{Gaia} satellite \citep{Brown2021b} and spectroscopic follow-up it has become evident that the Milky Way's halo is built from numerous substructures, providing clear signatures of both past and ongoing accretion events \citep{Belokurov2006, Myeong2018c, Koppelman2019, Naidu2020, Yuan2020, Dodd2023}. One particularly influential finding is that a major merger (hereafter called \textit{GSE}) with an estimated mass ratio between 1:2.5 and 1:4 \citep{Helmi2018, Naidu2020} occurred roughly 8–10~billion years ago, somewhat disrupting the protodisc and certainly contributing significantly to the halo. Early hints of this event came from stellar abundance patterns \citep{Nissen2010}, and were later confirmed with significant dynamical information from the \textit{Gaia} mission \citep{Brown2016, Brown2018} by \citet{Belokurov2018}, \citet{Haywood2018b}, and \citet{Helmi2018}, giving new weight to the proposed gas-rich merger hypothesis.

As summarised by \citet{Helmi2020}, the focus has now shifted from merely identifying accreted stars to characterising the properties of the merging progenitor galaxy and its gas content. Several studies have investigated its global chemical composition \citep[for example][]{Das2020, Feuillet2021, Aguado2021, Matsuno2021, Buder2022, Belokurov2022, DeSilva2023, Monty2024, Ou2024}, while others tried to recover the merger process itself \citep[for example][]{Naidu2021} as well as the effects of the merger on the existing stars, such as the heating -- or \textit{splashing} -- of stellar orbits \citep{DiMatteo2019, Belokurov2020, Belokurov2022}. More recently, \citet{Skuladottir2025}, building on N-body simulations by \citet{Mori2024}, examined how chemical properties vary with orbital characteristics. Even with fewer than one hundred stars in their updated sample\changed{, based on the high-resolution abundance measurements of \citet{Nissen2010} and \citet{Nissen2024},} they successfully separated the accreted stars into more and less tightly bound populations using orbital energies and identified statistically significant chemical differences between them. \citet{Skuladottir2025} interpreted these trends using the scenario proposed by \citet{Mori2024}, in which the outer stars of the accreted galaxy—characterised by higher (less negative) orbital energies—were accreted earlier and at larger Galactocentric radii, while stars from the central regions, with lower (more negative) energies, were accreted later and more centrally. The inferred radial gradient in star formation efficiency led to systematically higher $\mathrm{[Fe/H]}$ and $\mathrm{[Mg/Fe]}$ abundances in stars formed in the former core of the galaxy, a picture consistent with observations of present-day dwarf galaxies (Lucchesi et al. in prep). \changed{These results raise the possibility that present-day integrals of motion may preserve information about the internal chemical structure of the accreted progenitor galaxy, a scenario that can be tested directly in cosmological simulations that track stellar birth locations.}

To further test and refine this scenario, we turn to a different simulation suite: the NIHAO-UHD\thanks{NIHAO-UHD is the Ultra High Definition re-run of Milky Way analogues from the \textit{Numerical Investigation of a Hundred Astronomical Objects} (NIHAO) suite \citep{Wang2015}.} project, which provides high-resolution, fully cosmological zoom-in simulations of Milky Way-type galaxies undergoing satellite accretion \citep{Buck2020, Buck2020b, Buck2021}. Unlike the idealised N-body simulations by \citet{Mori2024}, NIHAO-UHD self-consistently models the chemical and dynamical evolution of stars within a hierarchical cosmological context \citep{Buck2021}. \changed{Similar to other cosmological zoom-in simulations of Milky Way analogues such as Auriga \citep{Grand2017} and the HESTIA suite \citep{Khoperskov2023}, the NIHAO-UHD simulations follow chemical enrichment from core-collapse supernovae, Type~Ia supernovae, and asymptotic giant branch stars self-consistently within a hierarchical galaxy assembly history \citep{Buck2021}. These implementations are less extensive in their treatment of individual nucleosynthetic channels than the detailed chemodynamical models of \citet{Kobayashi2011,Kobayashi2020}, which explicitly include metallicity-dependent yields from core-collapse supernovae and hypernovae, Type~Ia supernovae with metallicity-dependent delay-time distributions, asymptotic giant branch stars, and r-process enrichment from neutron-star mergers across a large number of elements. However, the cosmological zoom-in framework of NIHAO-UHD enables us to connect internal abundance gradients in an accreted progenitor galaxy directly to its merger history and present-day phase-space structure, which is essential for the analysis presented here.
} The simulation in particular also includes two post-processing steps that make it particularly well suited to study the chemodynamical memory of the galaxy and its major mergers. Firstly, we have traced the birth positions of stars in discrete $100\,\mathrm{Myr}$ time steps, enabling a clean separation of in-situ from accreted stars\footnote{Potential inaccuracies arise around the time of the major merger itself.}. Secondly, we have calculated the same integrals of motion, that is orbit\changed{al} energies and actions for the Milky Way analogue simulation that are calculated for stars in our Milky Way galaxy.

\changed{Similar analyses of accreted structures in cosmological simulations have been presented by \citet{Khoperskov2023d,Khoperskov2023b}, who examined the phase-space distribution of stellar-halo debris across the HESTIA suite of Local-Group analogues. Their work demonstrated that merger remnants occupy broad and evolving regions in energy--angular momentum and action space and already hinted at correlations between stellar age, chemistry, and binding energy. A complementary study by \citet{Khoperskov2023c} used these simulations to constrain the abundance gradient of the \textit{GSE} progenitor, establishing a connection between chemical and dynamical memory similar in spirit to the one explored here. Together with the trends identified observationally by \citet{Skuladottir2025} and in the Auriga simulations by \citet{Carrillo2026}, these results suggest that chemical abundance patterns in accreted stellar populations may retain measurable information about formation location within their progenitor galaxies. The NIHAO-UHD simulation analysed in this work allows us to test this emerging picture directly by tracing stellar birth positions and their subsequent evolution in integrals-of-motion space within a fully cosmological framework.}

In this first paper (Paper I) we address the following questions: 
\begin{enumerate}[leftmargin=2em,labelwidth=0em]
    \item What can we learn about the selection efficiency of accreted stars in integrals of motions phase-space \citep{Helmi2018, Feuillet2021, Buder2022, Monty2024}?
    \item What memory of the birth positions and past chemical evolution is retained in integrals of motion space \citep{Skuladottir2025}?
\end{enumerate}

Paper II \citep{Buder2025d} analyses how the effects of the merger process manifest in the simulation, for example through the proposed \textit{splashing} of in-situ star orbits \citep{Belokurov2020} or mixing of in-situ and accreted gas to form \textit{splashed} stars.

By answering these questions with a Milky Way analogue simulation, we reveal a preserved chemodynamical memory of the last major merger and unveil connections and biases that fundamentally change how we reconstruct the Milky Way's formative history. \changed{As we demonstrate below, this preserved chemodynamical memory provides a ``golden thread'' linking present-day chemistry and integrals-of-motion space to the internal formation structure of the accreted progenitor galaxy prior to its disruption.}

\section{Simulated data} \label{sec:data}

In this study we examine the high-resolution cosmological zoom-in simulation \texttt{g8.26e11}, a Milky Way analogue from the \textit{Numerical Investigation of a Hundred Astronomical Objects} (NIHAO) simulation suite \citep{Wang2015} that includes newly available chemical-evolution tracing \citep{Buck2021}.
We use \texttt{g8.26e11} because, as described below, it closely matches several key Milky Way properties — halo and stellar mass, rotation curve, and a disc-dominated, multi-armed spiral morphology — and it experienced a major merger at a lookback time comparable to the Milky Way's last major merger.
Although the model does not form a strong bar \citep[see][]{Buder2025}, \changed{this is not expected to significantly affect the conclusions of this work. Accreted stars are identified using their birth positions rather than present-day orbital properties, and radial migration driven by bars or spiral structure primarily changes angular momentum near corotation while largely preserving orbital eccentricity and radial action \citep{Sellwood2002,Solway2012}.}

This galaxy model forms part of the NIHAO-UHD series \citep{Buck2020b} and has been utilised in several prior investigations, including work on the properties of Milky Way satellites \citep{Buck2019b}, the spin of the Galactic dark matter halo \citep{Obreja2022}, abundance clustering to reconstruct birth conditions of stars \citep{Ratcliffe2022}, and the redshift evolution as well as shape of radial metallicity gradients in the ISM \citep{Ratcliffe2025, Buder2025}. A version of this simulation was also employed by \citet{Buck2023} and with reduced mass resolution by \citet{Buder2024} to explore the chemical tagging of stars associated with its last significant accretion event (\textit{GSE}).

\begin{figure}
    \centering
    \includegraphics[width=0.99\columnwidth]{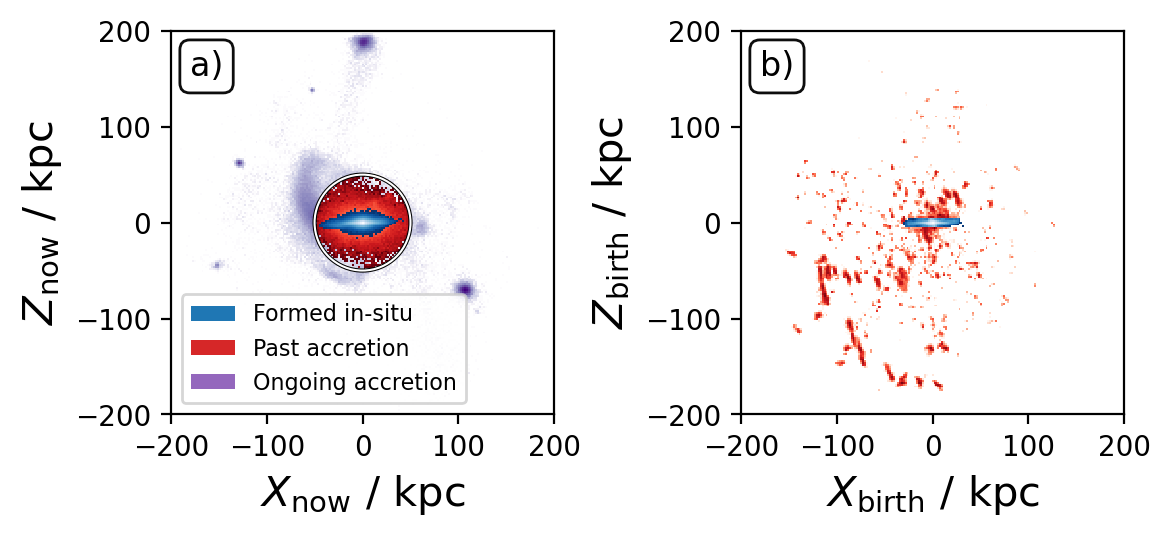}
    \caption{Tracing in-situ stars (blue) alongside past (red) and ongoing (purple) accretion components, shown in their present-day positions (panel~a) and birth (panel~b) positions (in 100 Myr intervals). The red overdensities in panel b primarily reflect the same accreted galaxy observed at different epochs \href{https://github.com/svenbuder/golden_thread_I/tree/main/figures}{\faGithub}.}
    \label{fig:tracing_insitu_accretion_2}
\end{figure}

We briefly describe the simulation with reference to the above mentioned works in Sec.~\ref{sec:data_simulation}. The main new insights for this simulation are provided by additional information on the birth positions of star particles (see Fig.~\ref{fig:tracing_insitu_accretion_2}) as well as orbit\changed{al} properties, which we describe in Secs.~\ref{sec:data_birth_positions} and \ref{sec:data_orbit_properties}, respectively.

\subsection{A NIHAO-UHD Milky Way analogue simulation} \label{sec:data_simulation}

The simulation was performed using the smoothed particle hydrodynamics code \texttt{Gasoline2} \citep{Wadsley2017}, incorporating sub-grid turbulent diffusion, and adopting cosmological parameters from \citet{Planck2014}. The simulation setup and feedback mechanisms follow the NIHAO framework \citep{Wang2015}, with the zoom-in configuration and baryonic physics described in \citet{Buck2021}. Star formation and stellar feedback prescriptions follow \citet{Stinson2006} and \citet{Stinson2013}, respectively. Notably, the simulation used here is a more recent, higher-resolution rerun of the model in \citet{Buder2024}, and includes updated nucleosynthetic yields.

We focus our analysis on the main halo by identifying it with the Amiga Halo Finder \citep{Knollman2009}, using the build-tin tools of the \textsc{pynbody} package \citep{pynbody}. To orient the system, we rotate it face-on by aligning the angular momentum vector via \textsc{pynbody.analysis.angmom.faceon}. The simulated disc galaxy with spiral arms has a virial radius of $R\_{200} = 206\,\mathrm{kpc}$ and a total mass (baryons + dark matter) within this radius of $9.1 \times 10^{11}\,\mathrm{M_\odot}$. At present-day ($z = 0$), this includes $8.2 \times 10^{11}\,\mathrm{M_\odot}$ in dark matter, $6.4 \times 10^{10}\,\mathrm{M_\odot}$ in gas, and $2.3 \times 10^{10}\,\mathrm{M_\odot}$ in stars. The stellar mass resolution is approximately $7.5 \times 10^3\,\mathrm{M_{\odot}}$ per particle. For this study, we will refer to the galactocentric radius as the three-dimensional spherical radius
\begin{equation}
    R_{\mathrm{3D}} = \sqrt{x^2 + y^2 + z^2},
\end{equation}
rather than the cylindrical one, which is often used when focusing on studies of galactic discs \citep[for example by][for the same simulation]{Buder2025}.

\begin{figure}
    \centering
    \includegraphics[width=\columnwidth]{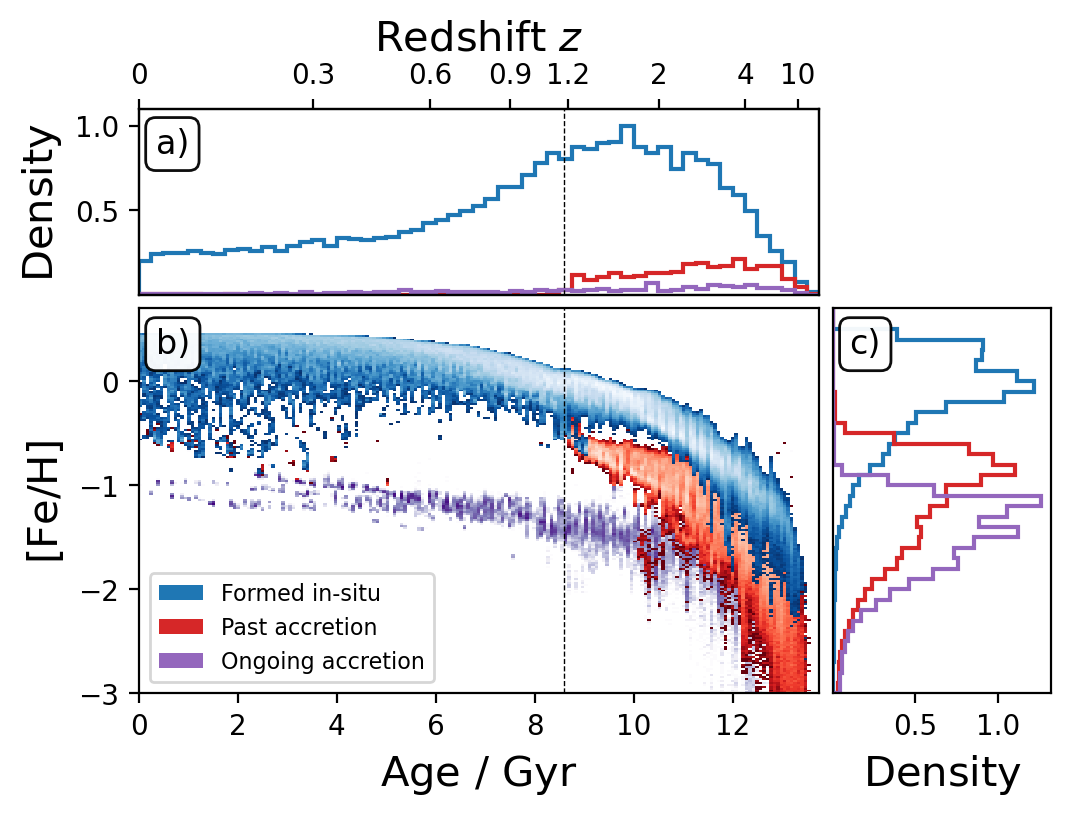}
    \caption{Age–metallicity distributions of in-situ (blue), previously accreted (red), and currently accreting (purple) stars, also shown as marginal 2D histograms with age (top) and [Fe/H] (right). A vertical dashed line indicates the time of the major merger around $8.6\,\mathrm{Gyr}$ ago \href{https://github.com/svenbuder/golden_thread_I/tree/main/figures}{\faGithub}.}
    \label{fig:tracing_insitu_accretion_3}
\end{figure}

Due to computational constraints, stars in the simulation are represented by tracer particles corresponding to simple stellar populations of uniform age, metallicity, and initial mass function (IMF). Chemical enrichment is computed using the \textsc{Chempy} framework \citep{Rybizki2017}, implemented by \citet{Buck2021}. \changed{In this implementation, \textsc{Chempy} is used to compute the time-dependent mass and element return of simple stellar populations (SSPs) as a function of metallicity, which are injected locally into the surrounding gas within the hydrodynamical simulation; the spatial distribution and mixing of metals are therefore governed self-consistently by the gas dynamics rather than by a one-zone chemical evolution model.} We adopt the \texttt{alt} configuration, assuming a \citet{Chabrier2003} IMF with a high-mass slope of $\alpha_\text{IMF} = -2.3$ , spanning stellar masses from 0.1 to $100\,\mathrm{M_\odot}$, and covering metallicities $Z/Z_\odot \in [10^{-5},2]$. The enrichment includes contributions from asymptotic giant branch (AGB) stars, core-collapse supernovae (CCSN; $8 - 40\,\mathrm{M_\odot}$), and Type Ia supernovae (SNIa), modelled with an exponential delay-time distribution (slope $-1.12$), a minimum delay of $40\,\mathrm{Myr}$, and a normalised rate of $\changed{\log_{10}}(N_\mathrm{Ia}) = -2.9$ (see \citealt{Buck2025,Guenes2025} for a Bayesian approach to selecting these parameters). Yields are taken from \citet[][CCSN]{Chieffi2004}, \citet[][SNIa]{Seitenzahl2013}, and \citet[][AGB; \texttt{new\_fid} yields in \citealt{Buck2021}]{Karakas2016}, but no additional contributions, for example from neutron star mergers, for the rapid neutron-capture process. The simulation tracks elemental abundances for H, He, C, N, O, Ne, Mg, Al, Si, P, S, V, Cr, Mn, Fe, Co, and Ba on a Solar abundance scale by \citet{Asplund2009}. In alignment with the approach in \citet{Buder2025} and differing from \citet{Buder2024}, we adopt the simulated abundances directly, without applying empirical offsets. We note, however, that the median $\mathrm{[Fe/H]}$ for stars within a Solar-like galactocentric cylindrical radius of $R_\mathrm{GC} = \sqrt{x^2+y^2} = 8.2 \pm 0.5\,\mathrm{kpc}$ \citep{BlandHawthorn_Gerhard2016} and Solar-like age of $4.5 \pm 0.5\,\mathrm{Gyr}$ \citep{Soderblom2010} is $+0.15\,\mathrm{dex}$ and thus slightly enhanced with respect to the actual Milky Way. The star formation history, approximated by the stellar age histogram in Fig.~\ref{fig:tracing_insitu_accretion_3}a, of our analogue and Milky Way estimates \citep{Snaith2015} is qualitatively similar with a steep rise from high redshifts to cosmic noon (at $z\sim2$) and a decreasing star formation thereafter. However, both differences in galaxy properties as well as our chosen yields and other chemical evolution parameters can explain the offsets in abundances \citep[see also][]{Buck2021}.

\subsection{Tracing birth positions for stars of the main galaxy body}  \label{sec:data_birth_positions}

In a post-processing step, we have selected all star particles within a radius of $50\,\mathrm{kpc}$ from the present-day centre of mass of the simulation box. We then went back to previous simulation snapshots and identified all star particles that were born between timesteps of $100\,\mathrm{Myr}$. We then logged the positions of these star particles relative to the centre of mass of the main halo at the time, effectively tracing the birth positions with respect to the main halo, including birth radii

\begin{equation}
    R_{\mathrm{birth}, \mathrm{3D}} = \sqrt{x_{\mathrm{birth}}^2 + y_{\mathrm{birth}}^2 + z_{\mathrm{birth}}^2}.
\end{equation}

\subsection{Orbital properties}  \label{sec:data_orbit_properties}

The simulation is tracing information on both the Cartesian velocities of each particle $v_x, v_y, v_z$ as well as the potential $\phi$ at its current position\footnote{This simulation output is \texttt{potential\_phi} in the simulation FITS file.}. This allows us to calculate the specific Newtonian energy $E$ of the particle (per unit mass) as
\begin{align}
    E = \frac{1}{2}\left( v_x^2 + v_y^2 + v_z^2 \right) + \phi(x,y,z).
\end{align}

In addition to the orbit\changed{al} energy, we also calculate orbit\changed{al} actions $J_{i \in [R, \varphi,z]}$ as integrals of motion with the \textsc{AGAMA} package \citep{Vasiliev2019b}. We construct the simulation potential component by component using \textsc{Potential}, representing the spheroidal halo with \changes{the spherical-harmonic \textsc{Multipole} expansion} and the flattened stellar and gaseous \changes{discs} with \changes{the azimuthal-harmonic \textsc{CylSpline} expansion, which is} better suited to disc geometries. 
The total potential is then defined as the sum of these individually fitted components. This composite potential is passed to the \textsc{ActionFinder}, which computes the approximate orbital actions ($J_R$, $J_\varphi \equiv L_Z$, and $J_Z$) for any given phase-space point (or in our case star particle) using the fast but approximating Stäckel fudge method \citep{Binney2012, Sanders2015b}.

\section{Analysis} \label{sec:analysis}

The ability to trace stellar birth positions in the NIHAO-UHD Milky Way analogue provides a unique opportunity to evaluate how much chemodynamical information is retained after a major merger. We begin by establishing a broad classification of stars into in-situ, past accretion, and ongoing accretion categories (Section~\ref{sec:analysis_broad_selection}). In Section~\ref{sec:analysis_dynamic_properties}, we examine the orbital properties of stars and quantify the efficiency of recovering accreted populations in integrals-of-motion space. Section~\ref{sec:analysis_chemodynamic_memory} then addresses how well the galaxy preserves its memory of stellar birth positions and chemical evolution in phase-space. Together, these analyses reveal how selection effects shape our ability to reconstruct the progenitor and chemical history of the last major merger.

\begin{figure*}
    \centering
    \includegraphics[width=\textwidth]{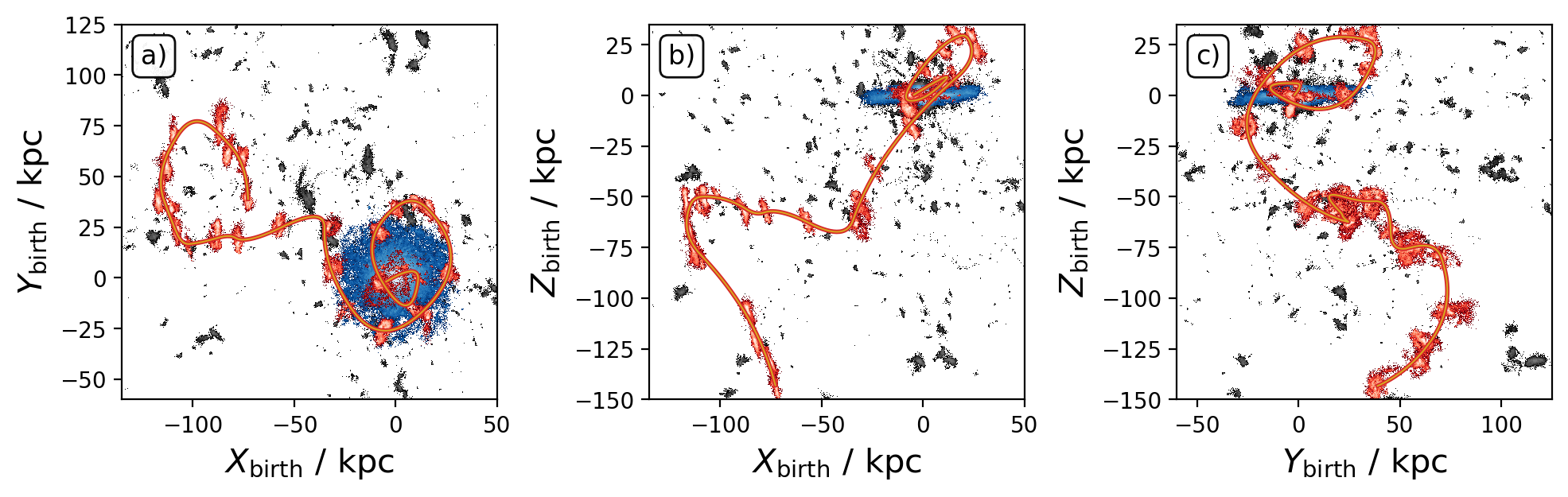}
    \caption{Birth positions in different Galactocentric planes of all star particles that are now within $50\,\mathrm{kpc}$ (grey density scale), where those born in-situ are in blue and those of the last major merger in red. Birth positions are estimated in $100\,\mathrm{Myr}$ steps and thus allow us to follow the changing position of star formation of the now accreted galaxy with respect to the Milky Way analogue. A dark golden line then interpolates the path of the now accreted galaxy \href{https://github.com/svenbuder/golden_thread_I/tree/main/figures}{\faGithub}.}
    \label{fig:tracing_xyz_birth_3}
\end{figure*}

\subsection{Tracing in-situ formation, past accretion, and ongoing accretion} \label{sec:analysis_broad_selection}

To broadly separate stars by their origin, we adopt a three-part classification: \textit{in-situ formation}, \textit{past accretion}, and \textit{ongoing accretion}. The goal is to construct a simple, interpretable, but still somewhat physically motivated scheme that captures the dominant contributors in each category without invoking overly complex selection criteria. Our classification is thus informed by both the present-day and birth positions of stars in the simulation (Fig.~\ref{fig:tracing_insitu_accretion_2}) as well as their age–metallicity distribution (Fig.~\ref{fig:tracing_insitu_accretion_3}). Based on these diagnostics, we define the following three selections. \changed{Each boundary is chosen to isolate visually distinct stellar populations identifiable either in present-day position space (Fig.~\ref{fig:tracing_insitu_accretion_2}) or in the age–metallicity relation (Fig.~\ref{fig:tracing_insitu_accretion_3}).}

Stars associated with \changed{currently} infalling dwarf galaxies \changed{form distinct young and metal-poor sequences in the age–metallicity relation (Fig.~\ref{fig:tracing_insitu_accretion_3}) and are either still located at large Galactocentric radii or have only recently become weakly bound during their ongoing accretion. Because tidal debris from such systems can already be partially captured, we do not rely on orbital energy alone but instead combine spatial, energetic, and chemical criteria to identify this population. These systems are selected using}
\begin{equation}
\text{Ongoing accretion:} \qquad
\begin{cases}
&R_\mathrm{3D} > 50\,\mathrm{kpc} \text{ or }\\
&E > 0\,\mathrm{kpc\,km\,s^{-1}} \text{ or } \\
&( \mathrm{age} < 10\,\mathrm{Gyr} \text{ and }  [\mathrm{Fe}/\mathrm{H}] < -1 )
\end{cases}
\end{equation}
\changed{The last adjustment ensures that younger metal-poor stars belonging to clearly separated dwarf galaxies visible as a distinct sequence in the age–metallicity relation in Fig.~\ref{fig:tracing_insitu_accretion_3} are not misclassified as part of the main galaxy or major accreted progenitor when their host galaxies have already crossed within $50\,\mathrm{kpc}$ and became bound.}

\changed{Stars associated with previously accreted systems are currently bound and located within $50\,\mathrm{kpc}$, but retain signatures of external origin in their birth positions, either through large birth radii or formation at significant vertical distances from the disc plane during the inclined merger phase. These stars are selected using}
\begin{equation} \label{eq:selection_past_accretion}
    \text{Past accretion:} \qquad \begin{cases}
    R_{\mathrm{3D}} \leq 50\,\mathrm{kpc} \text{ and }\\
    E < 0\,\mathrm{kpc\,km\,s^{-1}} \text{ and } \\
    ( R_{\mathrm{birth}, \mathrm{3D}} > 50\,\mathrm{kpc} \text{ or } |z_{\mathrm{birth}}| > 5\,\mathrm{kpc} )
    \end{cases}
\end{equation}
In particular the selection of stars with $|z_{\mathrm{birth}}| > 5\,\mathrm{kpc}$ is an important and clean adjustment, as it adds the $28\,\%$ of finally selected stars that were born in the last stages of the inclined merger. We have optimised this selection via the position of selected stars in the age–metallicity plane of Fig.~\ref{fig:tracing_insitu_accretion_2}b.

All remaining stars formed \changed{within the central disc region of the galaxy and confined to the disc plane at birth are classified as in-situ stars and selected using}
\begin{equation} \label{eq:selection_insitu}
\text{In-situ:} \qquad
\begin{cases} 
    R_{\mathrm{3D}} \leq 50\,\mathrm{kpc} \text{ and }\\
    E < 0\,\mathrm{kpc\,km\,s^{-1}} \text{ and } \\
    R_{\mathrm{birth}, \mathrm{3D}} \leq 50\,\mathrm{kpc} \text{ and }\\
    |z_{\mathrm{birth}}| \leq 5\,\mathrm{kpc}
\end{cases}
\end{equation}

While our selection is certainly not perfect, it provides a reasonable compromise. Each of the three groups is dominated by stars that plausibly reflect their intended origin, with minimal overlap. \changed{Although positive orbital energy is often used as a proxy for ongoing accretion, we do not rely on this criterion alone, since parts of satellites that are still in the process of being accreted can already be weakly bound to the main galaxy.} $95\%$ of the stars selected \changed{as part of the ongoing-accretion sample are nevertheless unbound at the present day, confirming that the selection predominantly identifies currently infalling systems while still retaining debris from satellites that are already partially captured and undergoing tidal disruption within $\sim 50\,\mathrm{kpc}$ (e.g.\ the system near $X \sim -50\,\mathrm{kpc}$ in Fig.~\ref{fig:tracing_insitu_accretion_2}).}

Similarly, $99\%$ of stars selected as accreted in the past are currently on bound orbits \changed{even before explicitly enforcing the negative orbital-energy criterion, confirming that this population is naturally associated with material already captured by the Milky Way analogue.} Tracing the star formation in Figs.~\ref{fig:tracing_insitu_accretion_2} and \ref{fig:tracing_insitu_accretion_3}, only 6771 star particles ($2.7\,\%$ of accreted sample) fall into a hard-to-classify category of overlapping chemistry ($-0.6 < \mathrm{[Fe/H]} < -0.2$) and birth positions between the last stage of the merger around $8.50$ and $8.65\,\mathrm{Gyr}$ ago. Our selection thus provides a useful basis for subsequent analysis.

Tracing the star formation in Figs.~\ref{fig:tracing_insitu_accretion_2} and \ref{fig:tracing_insitu_accretion_3}, only 6771 star particles ($2.7\,\%$ of accreted sample) fall into a hard-to-classify category of overlapping chemistry ($-0.6 < \mathrm{[Fe/H]} < -0.2$) and birth positions between the last stage of the merger around $8.50$ and $8.65\,\mathrm{Gyr}$ ago. Our selection thus provides a useful basis for subsequent analysis.

Of the $2.9 \times 10^6$ star particles ($2.3 \times 10^{10}\,\mathrm{M_\odot}$), we select $87.9\,\mathrm{\%}$ ($2.0 \times 10^{10}\,\mathrm{M_\odot}$) as in-situ, $8.5\,\mathrm{\%}$ ($1.9 \times 10^{9}\,\mathrm{M_\odot}$) as previously accreted and $3.5\,\mathrm{\%}$ ($8.1 \times 10^{8}\,\mathrm{M_\odot}$) as currently being accreted. At \changed{the present day,} the stellar mass ratio of previously accreted galaxy compared to the Milky Way analogue is 1:11. When only counting stars with ages above $8.6\,\mathrm{Gyr}$ ago, that is older than the merger, the mass ratio is 1:5 ($1.9 \times 10^{9}\,\mathrm{M_\odot}$ vs. $9.6 \times 10^{9}\,\mathrm{M_\odot}$).

While versions of Figs.~\ref{fig:tracing_insitu_accretion_2}a and \ref{fig:tracing_insitu_accretion_3} have already been analysed for a lower resolution version of this simulation by \citet{Buder2024}, the pattern of birth positions in Fig.~\ref{fig:tracing_insitu_accretion_2}b is intriguing, as we can see several overdensities that follow each other like a string of pearls.
For each of these pearls, representative of star formation within $100\,\mathrm{Myr}$ post-processing steps (see Sec.~\ref{sec:data_birth_positions}), we identify a spatial region within a reasonable age interval (see Tab.~\ref{tab:birth_position_tabular}) that traces the median birth position and then fit and interpolate a smooth function\footnote{We use \textsc{scipy.interpolate}'s \textsc{splprep} and \textsc{splev} functions \citep{Scipy} to fit and then evaluate a 2-degree B-spline function with smoothing value 100.} to the discrete positions.  

We visualise this in Fig.~\ref{fig:tracing_xyz_birth_3} as a trajectory that reliably\footnote{Due to the unstable orientation of the main galaxy for snapshots more than $12\,\mathrm{Gyr}$ ago, this reference frame is only reliable to trace the major merger galaxy (visible as several of the grey overdensities in Fig.~\ref{fig:tracing_xyz_birth_3}) after the main galaxy's orientation stabilised.} traces \changed{the sequence of median birth positions of the} merged galaxy based on the star formation from $12\,\mathrm{Gyr}$ ago until its last star formation period during the merger with the analogue around $8.6\,\mathrm{Gyr}$ ago \changed{in the progenitor-centred reference frame rather than reconstructing its orbital trajectory through the Milky Way halo.} While not the focus of our work, we note that the galaxy that later merged with the Milky Way encountered a merger of its own with another galaxy $10.8\,\mathrm{Gyr}$ ago around $(X_\mathrm{birth},Y_\mathrm{birth},Z_\mathrm{birth}) = (-120,25,-55)\,\mathrm{kpc}$ that changed its course. In almost all cases, we see the now accreted galaxy as \changed{a disc galaxy, most clearly in the nearly face-on projections across the time steps shown in Fig.~\ref{fig:tracing_xyz_birth_3}c.}

We will investigate how birth positions may be related to present-day properties in Sec.~\ref{sec:analysis_chemodynamic_memory}. But first, we provide an overview of orbit\changed{al} properties in the simulated galaxy.

\subsection{Dynamic properties and selection efficiency of accreted stars via integrals of motions} \label{sec:analysis_dynamic_properties}

\begin{figure*}
    \centering
    \includegraphics[width=\textwidth]{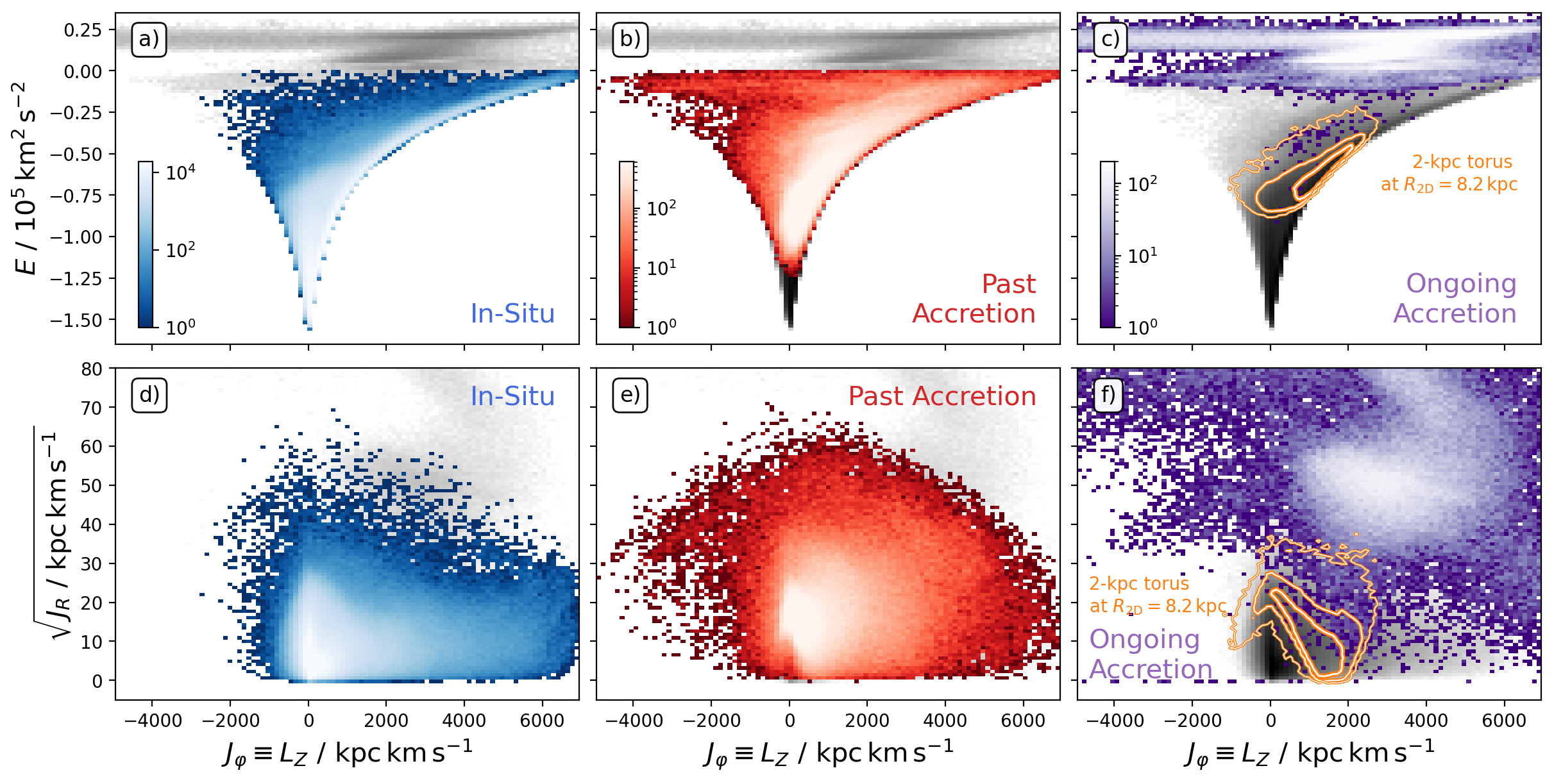}
    \caption{\changed{Orbital} properties of in-situ formed stars (blue, left panels), past accretion (red, middle panels), and ongoing accretion (purple, right panels), \changed{shown relative to the full stellar population (grey background in each panel).} Both the specific energy $E$ (upper panels) and the radial action $J_R$ (lower panels) are shown with the angular momentum $J_\varphi \equiv L_Z$. \changed{The concentration of ongoing-accretion stars at high energies reflects that these systems are weakly bound or unbound at the present day.} Stars undergoing accretion, currently located in the Solar-analogue neighbourhood of a 2\,kpc torus around $R_\mathrm{2D} = 8.2\,\mathrm{kpc}$, are indicated with orange density contours (the 68th, 95th, and 99.7th percentiles of stars) \href{https://github.com/svenbuder/golden_thread_I/tree/main/figures}{\faGithub}.}
    \label{fig:lz_e_jr}
\end{figure*}

As customary in recent observational Galactic dynamical studies \citep[for example][]{Helmi2018, Trick2019, Helmi2020, Buder2022}, we take a look at the distribution of stars in two of the integrals of motion, namely orbit\changed{al} energy $E$ over angular momentum $L_Z$ (top rows of Fig.~\ref{fig:lz_e_jr}) or a condensed version with radial action $J_R$ over angular momentum $L_Z$ (bottom rows of Fig.~\ref{fig:lz_e_jr} using $\sqrt{J_R}$).

Fig.~\ref{fig:lz_e_jr}a shows that most of the in-situ stars of the spiral galaxy are moving on circular disc orbits as they are concentrated along a curved ridge tracing the maximum angular momentum allowed at each energy. In Fig.~\ref{fig:lz_e_jr}d, these stars show low radial actions of $J_R < 10^2\,\mathrm{kpc\,km\,s^{-1}}$. In both figures, we note a broader excess of stars with higher (less negative) orbit\changed{al} energies ($E > -10^5\,\mathrm{km^2\,s^{-2}}$) and larger radial actions ($J_R \gtrsim 10^2\,\mathrm{kpc\,km\,s^{-1}}$) in the area of negligible net rotation ($L_Z \sim 0\,\mathrm{kpc\,km\,s^{-1}}$). This region of dynamically hotter stars with more (radially) eccentric orbits is also densely populated by stars that have been accreted previously (see Figs.~\ref{fig:lz_e_jr}b and \ref{fig:lz_e_jr}e around $L_Z \sim 0\,\mathrm{kpc\,km\,s^{-1}}$). The orbits of previously accreted stars extend from these regions of low absolute angular momentum $\vert L_Z\vert$ towards similar positive angular momenta, but show typically higher (less bound) orbit\changed{al} energies and larger radial actions, typical for more radial and eccentric orbits. Stars associated with ongoing accretion (Figs.~\ref{fig:lz_e_jr}c, \ref{fig:lz_e_jr}f), trace partially phase-mixed tidal debris from currently disrupting satellites as well as surviving satellites. They exhibit a wide range in angular momentum, often with retrograde orbits, and occur at high, predominantly positive energies.

As we show the whole galaxy in these figures, we note that the distributions in Fig.~\ref{fig:lz_e_jr} look different from what has been found observationally in the Milky Way's Solar neighbourhood \citep[see for example][]{Helmi2018,Trick2019,Das2020,Buder2022}. We have included orange contours in Figs.~\ref{fig:lz_e_jr}c and \ref{fig:lz_e_jr}f which trace the 68th, 95th, and 99.7th percentiles of stars in a torus with a tube radius of $2\,\mathrm{kpc}$ at a galactocentric cylindrical distance of $R_\mathrm{2D} = 8.2\,\mathrm{kpc}$, typical of footprints of current Milky Way surveys \citep[for example][]{SDSSDR17, Katz2023, Buder2025}. These are distributed mainly within angular momenta of $0 < L_Z~/~\mathrm{kpc\,km\,s^{-1}} < 2000$ and show circular orbits (low $J_R$) at their largest angular momenta and more eccentric, radial orbits for stars with low absolute angular momenta; in agreement with observations.

\begin{figure}
    \centering
    \includegraphics[width=0.94\linewidth]{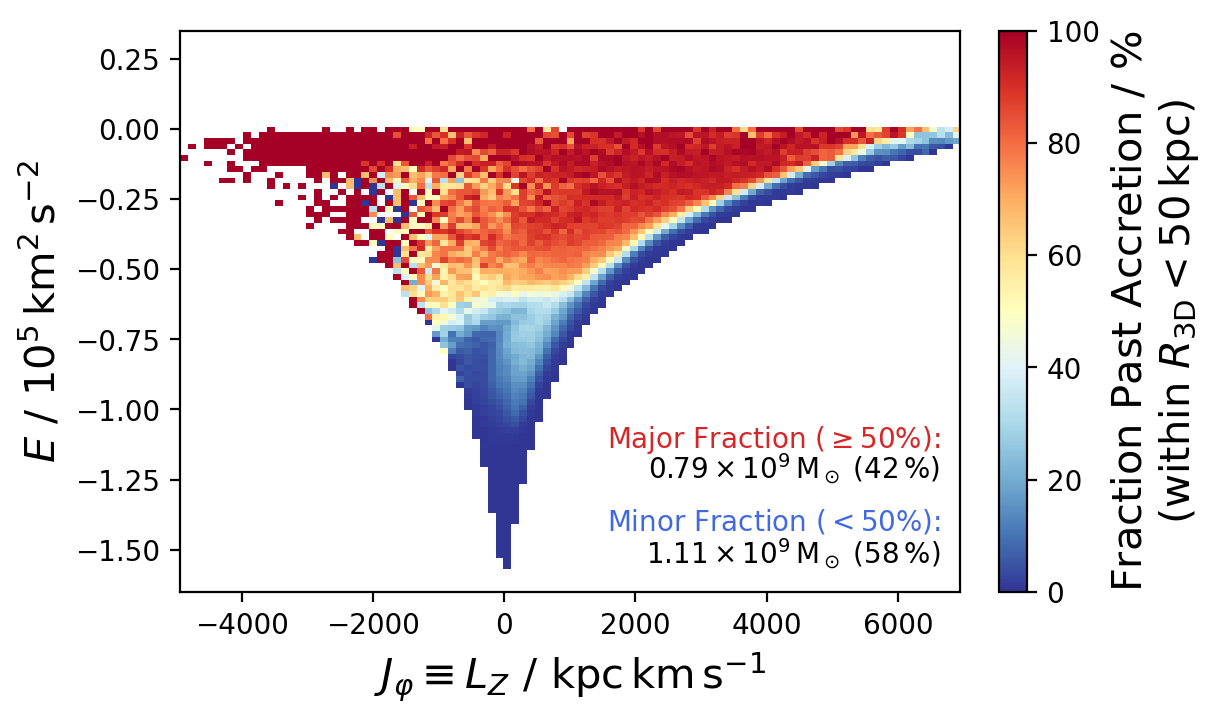}
    \includegraphics[width=0.94\linewidth]{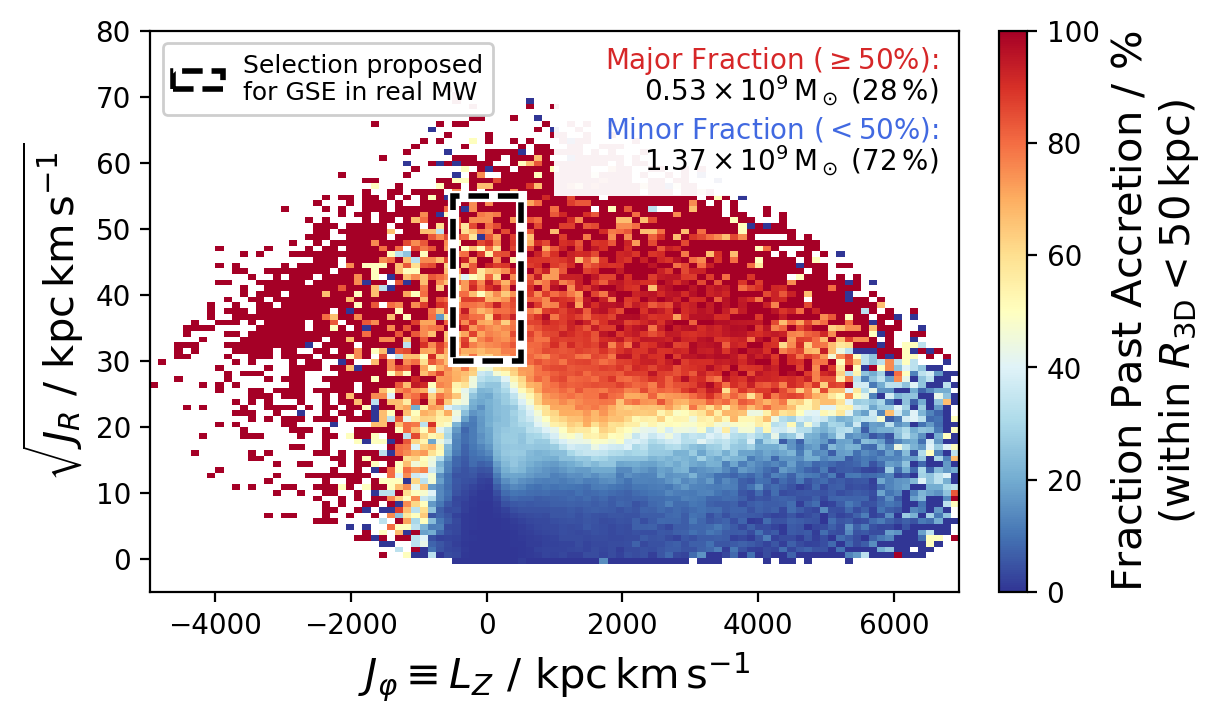}
    \caption{Angular momentum $J_\varphi \equiv L_Z$ vs.\ specific energy $E$ (top panel) and radial action $J_R$ (bottom panel) for stars within $R_\mathrm{3D} < 50\,\mathrm{kpc}$ at redshift $z=0$. \changed{Colours indicate, in each two-dimensional histogram cell (uniform binning over $-4950<L_Z<6950\,\mathrm{kpc\,km\,s^{-1}}$ with $\Delta L_Z\approx120\,\mathrm{kpc\,km\,s^{-1}}$, and either $-1.65<E<0.35\times10^5\,\mathrm{km^2\,s^{-2}}$ with $\Delta E\approx0.02\times10^5\,\mathrm{km^2\,s^{-2}}$ or $-5<\sqrt{J_R}<80\,\mathrm{kpc\,km\,s^{-1}}$ with $\Delta\sqrt{J_R}\approx0.86\,\mathrm{kpc\,km\,s^{-1}}$), the fraction of previously accreted stars relative to the total number of stars in that cell (defined as Figs.~\ref{fig:lz_e_jr}b/(a+b) and \ref{fig:lz_e_jr}d/(d+e), respectively).} We have added the selection proposed by \citet{Feuillet2020} for the real Milky Way as dashed black box. Fig.~\ref{fig:fraction_accreted_in_situ_lz_total} also includes ongoing accretion in the denominator \href{https://github.com/svenbuder/golden_thread_I/tree/main/figures}{\faGithub}.}
    \label{fig:fraction_accreted_in_situ_lz}
\end{figure}

In Fig.~\ref{fig:lz_e_jr}, we are especially intrigued by the extended distribution of accreted stars towards the lowest energies (Fig.~\ref{fig:lz_e_jr}b) and the significant overlap with the inner in-situ galaxy (Fig.~\ref{fig:lz_e_jr}a). These are regions that are not extensively explored by surveys of the Solar neighbourhood, but are expected to contain a significant fraction of accreted stars, well hidden within a labyrinth of in-situ stars. We have calculated the relative fraction of previously accreted stars to the main body of the galaxy ($R_\mathrm{3D} < 50\,\mathrm{kpc}$), in Fig.~\ref{fig:fraction_accreted_in_situ_lz}. Overall, we see that in-situ stars dominate (blue regions) for specific energies below $E < -0.58\times10^{5}\,\mathrm{kpc\,km\,s^{-1}}$ as well as radial actions below $J_R < 20^2\,\mathrm{kpc\,km\,s^{-1}}$ and in general stars on circular orbits, with the latter being situated along the \changed{rightmost} edge of the distribution. Accreted stars dominate (red regions) in the less bound and non-circular regions of the diagrams.

We note the remarkable similarity between Fig.~\ref{fig:fraction_accreted_in_situ_lz} and those presented by \citet[][their Fig.~1]{Belokurov2022} and \citet[][their Fig.~1]{Monty2024} who \changed{also} plot of angular momentum $J_\varphi = L_Z$ vs. specific energy $E$, but colour-coded by [Al/Fe] and [Eu/Si], respectively. The latter work found a significant change in the median chemical compositions around 50\% of the lowest orbit\changed{al} energies for their chosen potentials\footnote{The different amplitudes of the orbit\changed{al} energies that mark the boundary between being dominated by accreted or in-situ populations can be explained by both the different actual galaxies and potentials. We find $-0.66 \times 10^5\,\mathrm{km^2\,s^{-2}}$ in our simulation, while \citet{Monty2024} and \citet{Belokurov2022} find $-1.6 \times 10^5\,\mathrm{km^2\,s^{-2}}$ and $-0.75 \times 10^5\,\mathrm{km^2\,s^{-2}}$, respectively.} and attribute them to a transition of accreted to in-situ stars. We find a similar significant change of the dominant population from accreted to in-situ stars at about 50\,\% of the lowest orbital energy in our sample. As previously noted, we find that almost half the population of accreted stars is hidden in the region of $E < -0.58\times10^5\,\mathrm{km^2\,s^{-2}}$ that is dominated by in-situ stars. When selecting only from the regions in the $L_Z$-$E$ diagram with a fraction of past accretion larger than $50\mathrm{\%}$ (orange-red regions in Fig.~\ref{fig:fraction_accreted_in_situ_lz}), we only select $42\,\mathrm{\%}$%
 of stars of the last major merger.

For the simulated Milky Way analogue, we find that the accreted stars dominate in regions of more radial orbits, typically with $J_R > 20^2\,\mathrm{kpc\,km\,s^{-1}}$ and for the region of $\vert L_Z \vert \lesssim 1000\,\mathrm{kpc\,km\,s^{-1}}$ with slightly higher values of $J_R > 30^2\,\mathrm{kpc\,km\,s^{-1}}$ . As we are plotting the whole galaxy in this figure, we include a significant \changed{number} of stars with high radial actions, which are typically not visiting the Solar neighbourhood \citep[compare Figs.~\ref{fig:lz_e_jr}e and \ref{fig:lz_e_jr}f and see for example][]{Feuillet2019}. A significant region of the action-action space in the Milky Way is populated by other accreted systems \citep[for example][]{Myeong2019, Naidu2020}. It is thus not surprising, that the often suggested regions to cleanly select the stars of the last major merger in the Milky Way have been restricted to a small window around $\vert L_Z \vert < 500\,\mathrm{kpc\,km\,s^{-1}}$ and $30^2 < J_R~/~\mathrm{kpc\,km\,s^{-1}} < 55$ \citep{Feuillet2021, Buder2022}, which we have added to Fig.~\ref{fig:fraction_accreted_in_situ_lz}b for reference. Interestingly, our simulation indicates that the suggested lower limit in radial actions coincides with the region where the ratio of accreted to in-situ stars shifts. Contamination thus rapidly increases when accreted stars are selected solely by low radial actions in action space. \citet{Buder2022} found that only $29 \pm 1\,\mathrm{\%}$ of chemically identifiable accreted stars in their sample are within this clean dynamical selection box. Similarly, we find that a restricted selection from Fig.~\ref{fig:fraction_accreted_in_situ_lz} in regions where accreted stars dominate only selects $28\,\mathrm{\%}$%
 of all accreted stars \citep[for a study of other selection criteria and their purity as well as completeness see][]{Carrillo2024}.

\changed{Comparable overlap between accreted and in-situ stars in $E$–$L_Z$ has also been reported in the HESTIA simulations \citep{Khoperskov2023b}, where debris from major mergers spreads across a wide range of energies and angular momenta as the host potential evolves. Their results further showed that the apparent position of merger debris in integral-of-motion space can drift with cosmic time owing to the evolving gravitational potential, consistent with the broad, overlapping distributions seen in our NIHAO-UHD galaxy.}

For completeness, we have also calculated the percentage ratio of previously accreted stars to all stars in each of the $L_Z$-$E$ and $L_Z$-$\sqrt{J_R}$ bins. We append them to this paper in Fig.~\ref{fig:fraction_accreted_in_situ_lz_total} to visualise the relatively minor contaminating effect of ongoing accretion at higher energies, similar to that of the Sagittarius dwarf spheroidal galaxy in the Milky Way. However, the overall estimate of selecting accreted stars from dominant regions does not shift significantly, that is, only from $42\,\mathrm{\%}$%
 to $41\,\mathrm{\%}$%
. Similarly, including ongoing accretion to calculate the fraction in the $L_Z$ vs. $\sqrt{J_R}$ domain (compare Figs.~\ref{fig:fraction_accreted_in_situ_lz} and \ref{fig:fraction_accreted_in_situ_lz_total}) insignificantly changed the selection ratio from  to $26\,\mathrm{\%}$%
. As we are able to use the birth positions for a rather clean selection of in-situ and accreted stars, we can now turn to the analysis of the memory of the progenitor galaxy that might still be imprinted in the present-day properties of the accreted stars.

\subsection{\changed{Orbital memory}}
\label{sec:analysis_chemodynamic_memory}

\begin{figure}
    \centering
    \includegraphics[width=0.95\columnwidth]{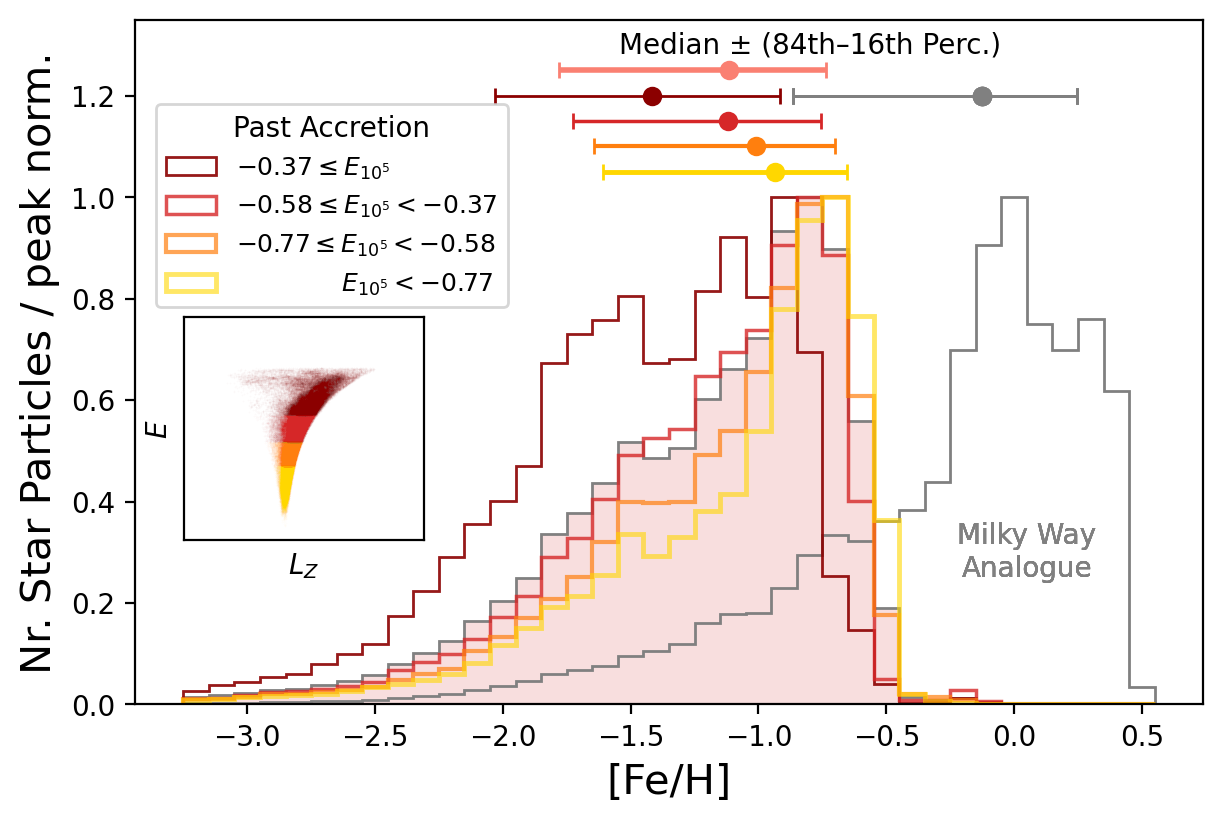}
    \caption{Histograms of ${\mathrm{[Fe/H]}}$ distributions of the whole galaxy (grey), and its previously accreted stars with different orbit\changed{al} energies as given in Table \ref{tab:energy_selection} (more negative energies from red to yellow, as shown in the inset panel of $L_Z$ vs. $E$). The percentiles of each distribution are plotted at the top right of the figure and show an increase of average [Fe/H] with more negative energies. See \href{https://github.com/svenbuder/golden_thread_I/tree/main/figures}{\faGithub} for individual figures showing the build-up of this plot.}
    \label{fig:fe_h_histograms}
\end{figure}

After presenting an overview of the energies and actions of in-situ and, more importantly, accreted stars in the simulation, we focus on how the properties of accreted stars depend on orbital energy $E$. This question was already investigated by \citet{Skuladottir2025}, using a small sample of 33 accreted stars with high-quality chemical abundance measurements, which they divide into two samples of 20 and 13 stars with energies above or below $E = -0.45\times10^{5}\,\mathrm{kpc\,km\,s^{-1}}$, respectively. Our selection of accreted stars consists of a quarter of a million star particles, enabling us to dissect our simulated galaxy into more samples. We thus decide to divide the sample into quartiles, or `zones', of orbit\changed{al} energy, which we list in Table~\ref{tab:energy_selection} and visualise through the inset shown in Fig.~\ref{fig:fe_h_histograms}.

\begin{table}
    \centering
    \renewcommand{\arraystretch}{1.25}
    \caption{Iron abundances [Fe/H] and stellar masses $M_\bigstar$ for the four energy-selected zones of previously accreted stars defined in Sec.~\ref{sec:analysis_chemodynamic_memory}. Median [Fe/H] values with $16^\mathrm{th}$–$84^\mathrm{th}$ percentiles are shown in Fig.~\ref{fig:fe_h_histograms}. Three logarithmic stellar masses (discussed in Sec.~\ref{sec:discussion_strategy_finding_gse_members}) are indicated: the total mass of the selected star particles ($\Sigma$), and the inferred masses from the reversed mass–metallicity relations of \citet[][Eq.~\ref{eq:naidu}]{Naidu2022b} and \citet[][Eq.~\ref{eq:kirby}]{Kirby2013} \href{https://github.com/svenbuder/golden_thread_I/tree/main/figures}{\faGithub}.}
    \begin{tabular}{ccccc}
\hline\hline
Energy $E$ & {[Fe/H]} & $M_\star (\Sigma)$ & $M_\star$ (Eq.~\ref{eq:naidu}) & $M_\star$ (Eq.~\ref{eq:kirby}) \\
$\mathrm{km^2\,s^{-2}}$ & dex & $\log_{10}\mathrm{M_\odot}$ & $\log_{10}\mathrm{M_\odot}$ & $\log_{10}\mathrm{M_\odot}$ \\
\hline
$-0.37 \leq$    & $-1.4_{-0.6}^{+0.5}$%
 & 8.7 & 7.9 & 7.9 \\
$[-0.58,-0.37]$ & $-1.1_{-0.6}^{+0.4}$%
 & 8.7 & 8.8 & 8.9 \\
$[-0.77,-0.58]$ & $-1.0_{-0.6}^{+0.3}$%
 & 8.7 & 9.1 & 9.3 \\
$<-0.77$        & $-0.9_{-0.7}^{+0.3}$%
 & 8.7 & 9.3 & 9.5 \\
\hline
All energies & $-1.1_{-0.7}^{+0.4}$%
 & 9.3 & 8.8 & 8.9 \\
\hline\hline
\end{tabular}

    \label{tab:energy_selection}
    \renewcommand{\arraystretch}{1.0}
\end{table}

These four samples, selected to have roughly 25\% of accreted stars in each of them, now allow us to investigate the memory or different properties across different orbit\changed{al} energies. In particular, we are interested in the change of the (i)~overall metallicity [Fe/H], (ii)~chemical abundances, (iii)~present-day positions, and (iv)~birth positions, which we analyse separately below.

\begin{figure*}
    \centering
    \includegraphics[width=0.6\textwidth]{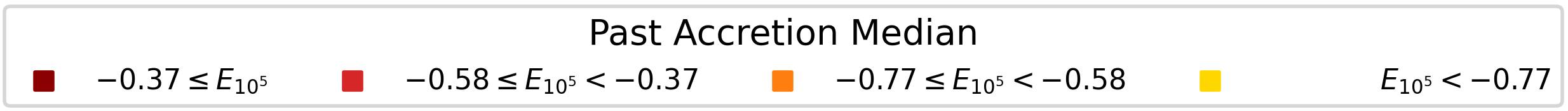}\\
    \includegraphics[width=0.33\textwidth]{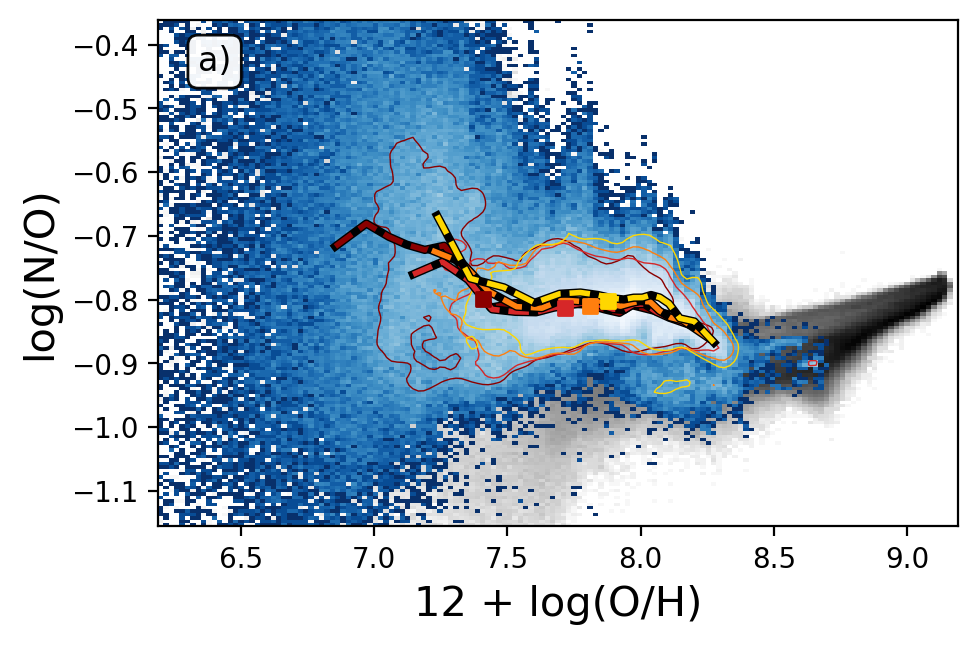}    \includegraphics[width=0.33\textwidth]{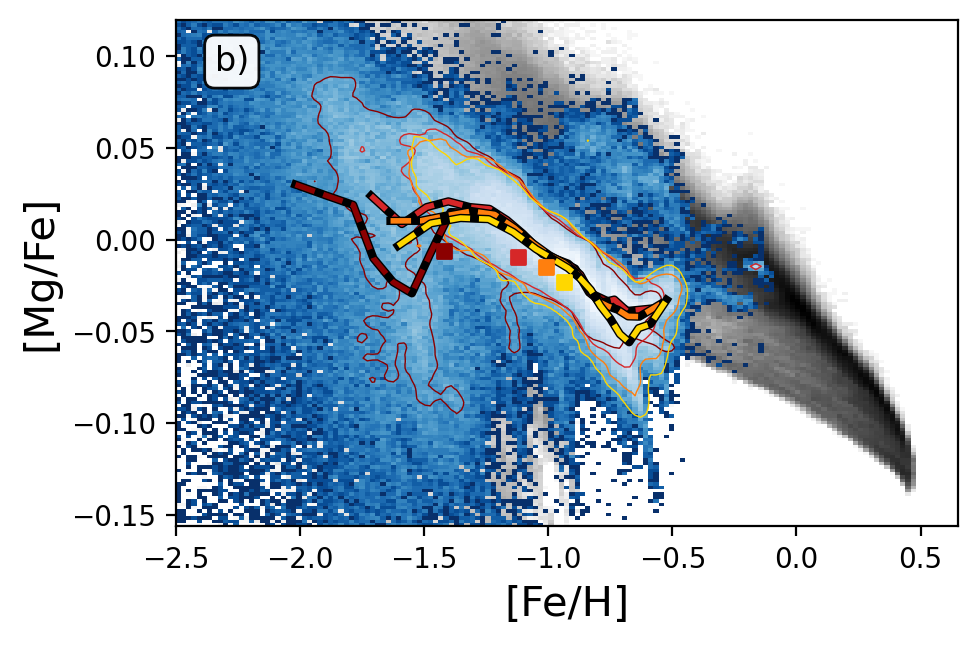}
    \includegraphics[width=0.33\textwidth]{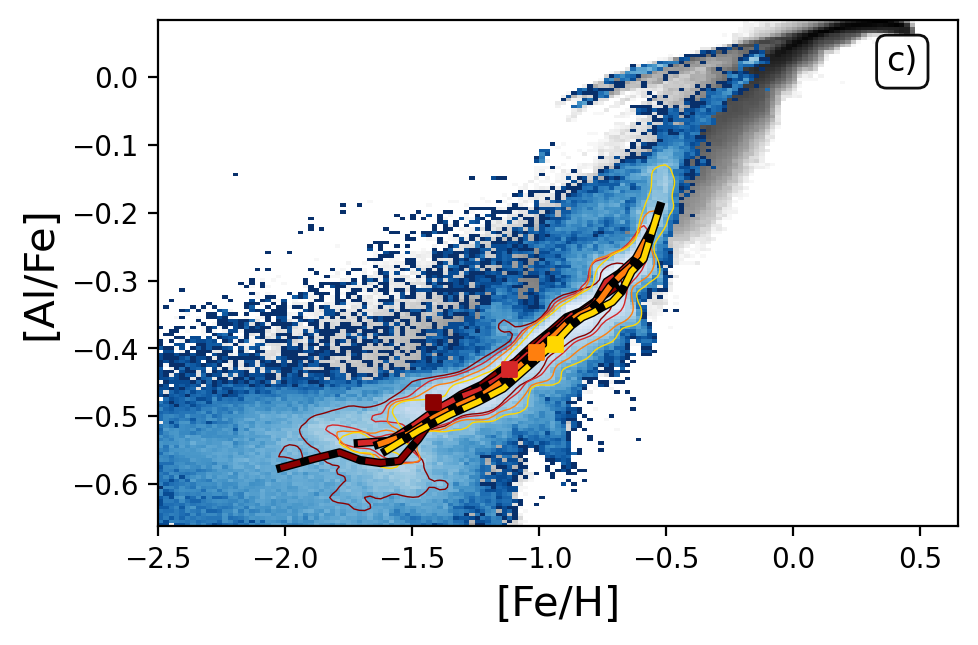}
    \includegraphics[width=0.33\textwidth]{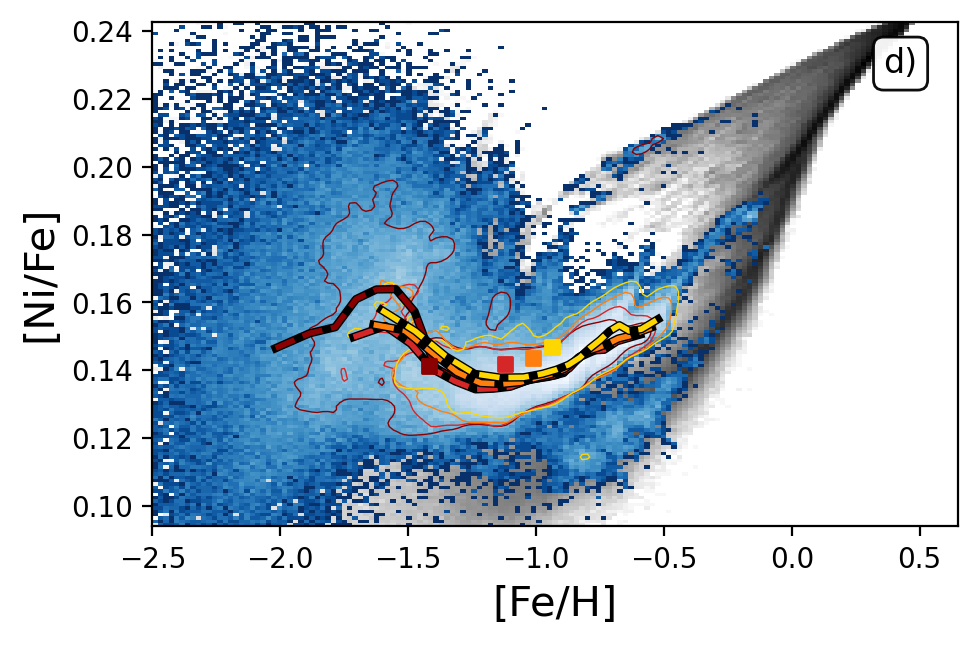}
    \includegraphics[width=0.33\textwidth]{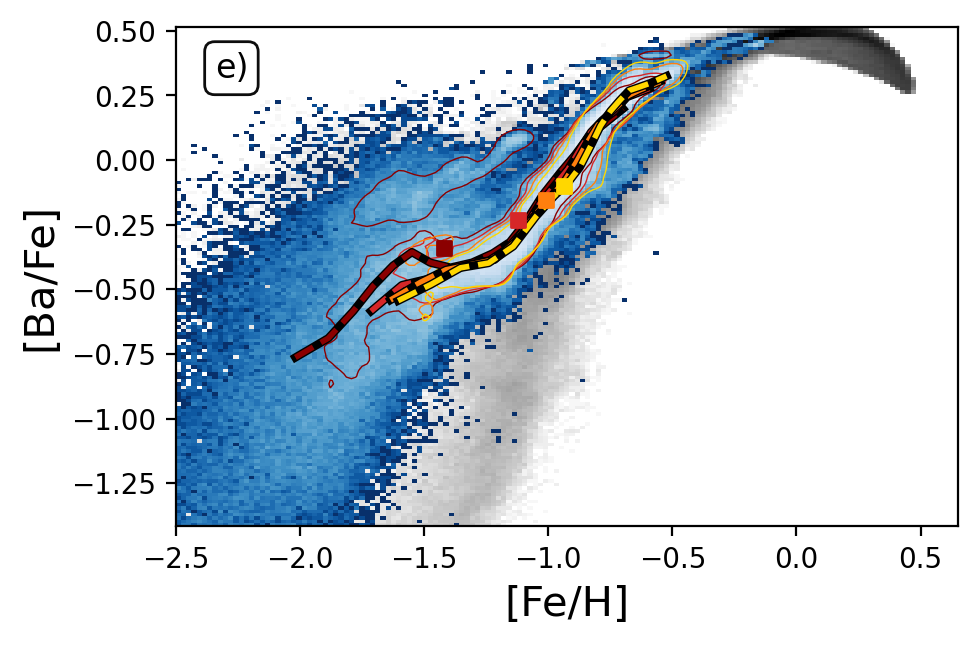}
    \includegraphics[width=0.33\textwidth]{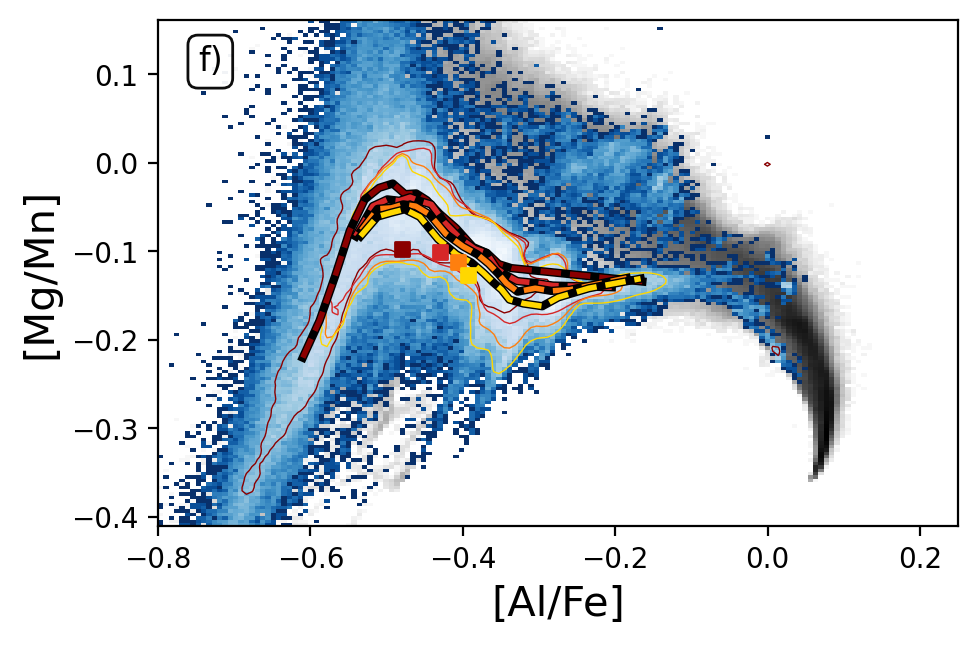}
        \caption{Abundance distributions of (past) accreted stars in blue, while the whole simulation is shown in the background (black). Yellow, orange, light and dark red dashed lines show the median abundances for 5-percent-bins if the x-axis of accreted stars for the different energy quantiles (same as in Fig.~\ref{fig:fe_h_histograms}). Narrow contours show the distribution of 68\% of each sample. Squares show the median abundances of each sample \href{https://github.com/svenbuder/golden_thread_I/tree/main/figures}{\faGithub}.}
    \label{fig:xfe_feh_zones}
\end{figure*}

\subsubsection{Orbital memory of {[Fe/H]} abundance}

In Fig.~\ref{fig:fe_h_histograms}, we plot the histograms of iron abundance [Fe/H] \changed{for accreted stars in the different energy zones} with different colours (yellow to dark red for increasing $E$) and the normalised histogram of all accreted stars as filled histogram in the background. The median values of accreted stars ($\mathrm{[Fe/H]} = $$-1.1_{-0.7}^{+0.4}$%
) are 1\,dex lower than the in-situ galaxy, which has a median $\mathrm{[Fe/H]} = $$-0.1_{-0.5}^{+0.3}$%
, where the errors are the $16-84^\mathrm{th}$ percentiles. This is consistent with expectations from the mass-metallicity relation \citep[for example][]{Gallazzi2005, Kirby2013} and the lower mass of the accreted system in comparison to the main galaxy (1:11).

For most distributions, we see asymmetric, negatively skewed distributions in logarithmic [Fe/H] abundance space, which constitute normal distributions in linear $N_\mathrm{Fe}/N_\mathrm{H}$ number density space. Compared to the other zones, the highest orbit\changed{al} energies (dark red) shows a broader and double-peaked distribution. This region is particularly prone to contamination from the many smaller accretion events of metal-poor systems (visible as many smaller dots in Fig.~\ref{fig:tracing_insitu_accretion_2}b). When comparing our overall selection of accreted stars in this energy zone with those that follow the trajectory of Fig.~\ref{fig:tracing_xyz_birth_3}, we find a contamination of up to 50\,\% for this energy zone. Such a contamination could be limited in the simulation when limiting the selection of the major merger with the birth positions or when using chemical selections in which the different star formation histories of small systems and the major merger galaxy manifest in different enhancement pattern. In observations, these additional aids would, however, either not be available or be observationally expensive. Furthermore, the [Fe/H] distribution of this energy zone, even at its upper end, is lower than those of the other energy zones.

Even when neglecting this zone, we note a decreasing trend of median [Fe/H] from $-1.1_{-0.6}^{+0.4}$%
 to $-0.9_{-0.7}^{+0.3}$%
 with increasing orbit\changed{al} energy. Both the decrease in [Fe/H] and the non-Gaussian distribution of [Fe/H] for accreted stars with the highest orbit\changed{al} energies agree with the observational findings by \citet[][see their Fig.~3]{Skuladottir2025} for the major accretion event in the Milky Way. \changed{A similar negative correlation between orbital energy and metallicity is seen in the HESTIA simulations by \citet[][see their Figs.~8 and~9]{Khoperskov2023c}. In these simulations, the most metal-rich stars of accreted galaxies occupy the lowest energies after merging, while the more metal-poor outskirts populate higher-energy orbits.}

\subsubsection{Orbit\changed{al} memory of abundances ratios}

Having established an energy-metallicity correlation, we are now concerned with the behaviour of elemental abundances beyond [Fe/H]. After having plotted all available abundances\footnote{We attach the remaining elements in Fig.~\ref{fig:additional_xfe_feh_zones}, as these either behave similar to the selected elements or might be unreliable due to our limited understanding of their modelling.}, we focus on six of the most insightful abundance combinations in Fig.~\ref{fig:xfe_feh_zones}, that is, the abundance planes of $\log(\mathrm{O/H}) + 12$ vs. $\log(\mathrm{N/O})$ in Fig.~\ref{fig:xfe_feh_zones}a, [Fe/H] vs. [Mg/Fe], [Al/Fe], [Ni/Fe], and [Ba/Fe] in Figs.~\ref{fig:xfe_feh_zones}b-e and the observationally found diagnostic plane of accreted stars of [Al/Fe] vs. [Mg/Mn] \citep{Hawkins2015, Das2020} in Fig.~\ref{fig:xfe_feh_zones}f. In this figure, we trace the running median abundances of the y-axis as dashed lines across 17 equal-number quantile bins\footnote{To limit the influence of contamination on the trend estimates, we limit the range to the $14^\mathrm{th}..5..99^\mathrm{th}$ percentiles.} in the x-axis. We utilise this method, because neither the median abundance in each dimension, shown as squares in the figure, nor linear functions as used by \citet{Skuladottir2025}, accurately captured the often non-linear abundance distributions. To grasp the exten\changed{t} of abundances covered by each sample, we also show the 68\% contours of the different energy samples as narrow lines on top of the density distribution of all accreted stars (blue) and the whole simulation (grey).

For $\log(\mathrm{N/O})$ as part of the CNO elements in Fig.~\ref{fig:xfe_feh_zones}a, we find that the overall downwards trends of [N/Fe] and [O/Fe] with increasing [Fe/H] lead to a rather flat relation with $\log(\mathrm{N/O}) \approx -0.8$. \changed{However, accreted stars in the highest-energy bin (dark red) show systematically enhanced $\log(\mathrm{N/O}) > -0.7$ at the lowest oxygen abundances of $12 + \log(\mathrm{O/H}) < 7.5$ ($\mathrm{[Fe/H]} < -1.5$), compared to the other energy zones.} These stars, with typical ages of $11.7_{-1.4}^{+1.1}\,\mathrm{Gyr}$ (corresponding to formation redshifts of $z = 3.0_{-1.1}^{+2.5}$), also exhibit higher [N/Fe] and weaker enhancement in the $\alpha$-process abundances (Fig.~\ref{fig:xfe_feh_zones}b and Fig.~\ref{fig:additional_xfe_feh_zones} for [O/Fe]). \changed{Together, these trends lead to elevated $\log(\mathrm{N/O})$ ratios at low metallicity that are reminiscent of, although not as extreme as, those observed in some high-redshift galaxies with JWST \citep{Cameron2023, Senchyna2024, Ji2026}.} \changed{Across the full energy range, however, the median variation remains small, with $\vert\Delta\log(\mathrm{N/O})\vert < 0.03$.} Neither N nor O were analysed by \citet{Skuladottir2025}.

For Mg, one of the $\upalpha$-process elements, [Mg/Fe] decreases mostly linearly with increasing [Fe/H] among the accreted stars in Fig.~\ref{fig:xfe_feh_zones}b, consistent with the onset of SNIa. In addition to low [Mg/Fe] stars in the highest energy group, we also note a slight upturn or spread of [Mg/Fe] of 0.03\,dex at the highest $\mathrm{[Fe/H]} \sim -0.5$. Contrary to \citet{Skuladottir2025}, we find very similar median abundance trends without a slope difference (compare to their Fig.~3)\changed{, consistent with similar abundance trends of accreted systems for $\mathrm{[Fe/H] < -1}$ in the HESTIA simulations \citep{Khoperskov2023c}}. We do find, however significantly different trends for the highest $\mathrm{[Fe/H]} > -0.75$, where we see both an increase in abundance spread, as well as an upturn of median abundances. This trend is consistent with the findings by \citet{Skuladottir2025} for their lowest energy sample (compare to their Fig.~5). We notice a similar behaviour of higher abundances for stars with lower orbit\changed{al} energies at similar [Fe/H] for the other $\upalpha$-process elements, such as Si -- again consistent with \citet{Skuladottir2025} -- as well as O, Ne, S, and Ti (Fig.~\ref{fig:additional_xfe_feh_zones}). Because the simulation did not include Ca, we cannot compare the trends with those by \citet{Skuladottir2025}, who find less of an increase or spread for this element.

For Al, one of the odd-Z elements, we find a steep increase of [Al/Fe] with rising [Fe/H] of almost $0.3\,\mathrm{dex}$ (from $-0.5$ to $-0.2$), while always remaining sub-solar in Fig.~\ref{fig:xfe_feh_zones}c. This \changed{trend results in systematically} higher [Al/Fe] values for \changed{stars with lower orbital energies} (see median values shown as squares). While the on average higher [Al/Fe] for stars with lower orbital energies agrees with \citet{Skuladottir2025}, they find a flat trend of [Al/Fe] with increasing [Fe/H] \citep[see also][]{Feuillet2021, Ernandes2025}. However, our trends of a steep increase of abundance agree for Na as another odd-Z element analysed by \citet[][see their Fig.~6]{Skuladottir2025}, when neglecting the two significant outliers with $\mathrm{[Na/Fe]} > 0$. Albeit not as strong, \citet{Belokurov2022} also find a slight increase in their accreted populations (see their Fig.~2). \changed{Since Al production includes both primary and metallicity-dependent secondary contributions from massive stars \citep[e.g.][]{Kobayashi2020}, an increase of [Al/Fe] toward lower orbital energies is consistent with stars originating in regions of higher star-formation efficiency and chemical enrichment within the progenitor galaxy.}

\begin{figure*}
    \centering
    \includegraphics[width=\textwidth]{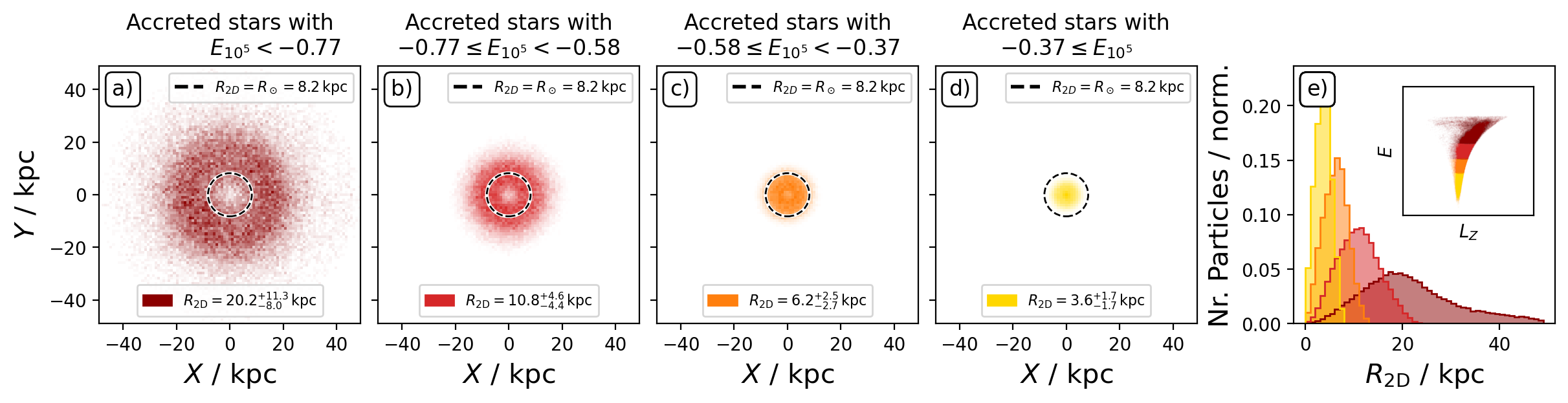}
    \caption{Present-day spatial distribution of accreted stars in the Galactocentric $X$-$Y$ plane. Panels a-d) show the density distribution of accreted stars with highest to lowest orbit\changed{al} energy quartiles, where the dashed circle shows a solar-analogue $R_\mathrm{2D}=8.2$\,kpc. Panel legends list the 50th and 16th-84th percentiles of galactocentric cylindrical radii $R_\mathrm{2D}$. Panel e) shows the $R_\mathrm{2D}$ distributions, with an inset visualising the $L_Z$ vs. $E$ selection. See Fig.~\ref{fig:xz_distribution_ezones} for a similar figure showing the $X$-$Z$ plane \href{https://github.com/svenbuder/golden_thread_I/tree/main/figures}{\faGithub}.}
    \label{fig:xy_distribution_ezones}
\end{figure*}

For Ni, one of the iron-peak elements, we find a rather flat trend of [Ni/Fe] with a parabola-like increase of 0.02\,dex in either [Fe/H] direction that agrees among the different energy samples. However, due to the different [Fe/H] distributions of the energy samples, this leads to an on average higher [Ni/Fe] with decreasing orbit\changed{al} energy. While \citet{Skuladottir2025} found an overall slightly decreasing linear trend for [Ni/Fe], they also find that the [Ni/Fe] is higher for lower orbit\changed{al} energies at the highest [Fe/H], consistent with the parabola-like shape present in the simulation.

For Ba, one of the neutron-capture elements, we find no clear agreement from the chemical evolution modelling in the simulation to the observational trends of the last major merger in the Milky Way when comparing with the default separation by \citet[][their Fig.~6]{Skuladottir2025}. We find that Ba, as well as Y and the other neutron-capture elements of the simulation, show a steep and almost linear increase with increasing [Fe/H]\changed{, in agreement with predictions by the HESTIA simulations for accreted systems \citep{Khoperskov2023c}}. \citet{Skuladottir2025} find a decreasing trends for [Y/Fe] in the observational data and an even more complicated scattered picture for [Ba/Fe]. We note though, that these disagreements can be reconciled for [Ba/Fe], when treating the three most metal-poor stars on low orbit\changed{al} energies as outliers in Fig.~6 of \citet{Skuladottir2025}. In that case, their trend of [Ba/Fe] could be interpreted as a steep and linearly inclining trend with increasing [Fe/H] and decreasing orbit\changed{al} energy -- now consistent with the simulation predictions. This is remarkable, as the nucleosynthesis tracing of neutron-capture elements is incomplete in the simulation due to the missing prescriptions for enhancement channels beyond CC-SN, SNIa, and the contribution from AGB stars.

For the often used diagnostic plot of accreted stars, [Al/Fe] vs. [Mg/Mn], we find a trend of decreasing [Mg/Mn] and increasing [Al/Fe] for the decreasing orbit\changed{al} energies. In the simulation, these trends are driven by the difference in Al and Mg, respectively, since Mn is showing a less pronounced change. We note though, that the combination of [Mg/Mn] leads to a slightly more pronounced energy-dependent spread \changed{in} median abundances of up to $\vert\Delta\mathrm{[Mg/Mn]}\vert = 0.05$ between the highest and lowest energy zones across the [Al/Fe] range.

Across all figures we find that the median abundance trends between different energy levels are rather similar and less pronounced than the observational trends found by \citet{Skuladottir2025}. Secondly, we find that the abundances tend to not exactly follow a linear trend, as fitted by \citet{Skuladottir2025}, but a more complex, non-linear behaviour \changed{in agreement with predictions by \citet{Khoperskov2023c}}. In combination with the different [Fe/H] distributions of the different energy samples, this leads to noticeable differences in the median abundances. Especially when comparing with the abundance trends for the lower energy cuts by \citet[][their Fig.~5]{Skuladottir2025}, we find very good agreement in the relative abundance changes between their lower and higher energy samples, in particular for the highest [Fe/H] regime.

As the simulation necessarily relies on specific choices for nucleosynthesis channels and stellar yields, we note that both the absolute abundance differences between accreted and in-situ populations and the intrinsic abundance spreads within each population are significantly smaller than observed in the Milky Way. \changed{While offsets in the absolute abundance scale are primarily driven by the adopted stellar yields, the reduced intrinsic abundance spreads in the simulation more likely reflect efficient metal mixing and the representation of stellar populations as single-age, single-metallicity particles, which suppress stochastic enrichment effects present in real galaxies \citep[see also discussion in][]{Buck2021}. In addition, part of the larger abundance scatter observed in the Milky Way reflects observational uncertainties and survey systematics, which are not present in the simulation.} This highlights a general caveat of our analysis: while the relative trends are informative, the absolute abundance scales and dispersions should not be over-interpreted \changed{when comparing to observations}, as they depend on both the adopted yields and the chemical-enrichment implementation \citep[see][for machine-learning-based model comparison approaches to select chemical enrichment models]{Buck2025,Guenes2025}.

\begin{figure}
    \centering
    \includegraphics[width=\columnwidth]{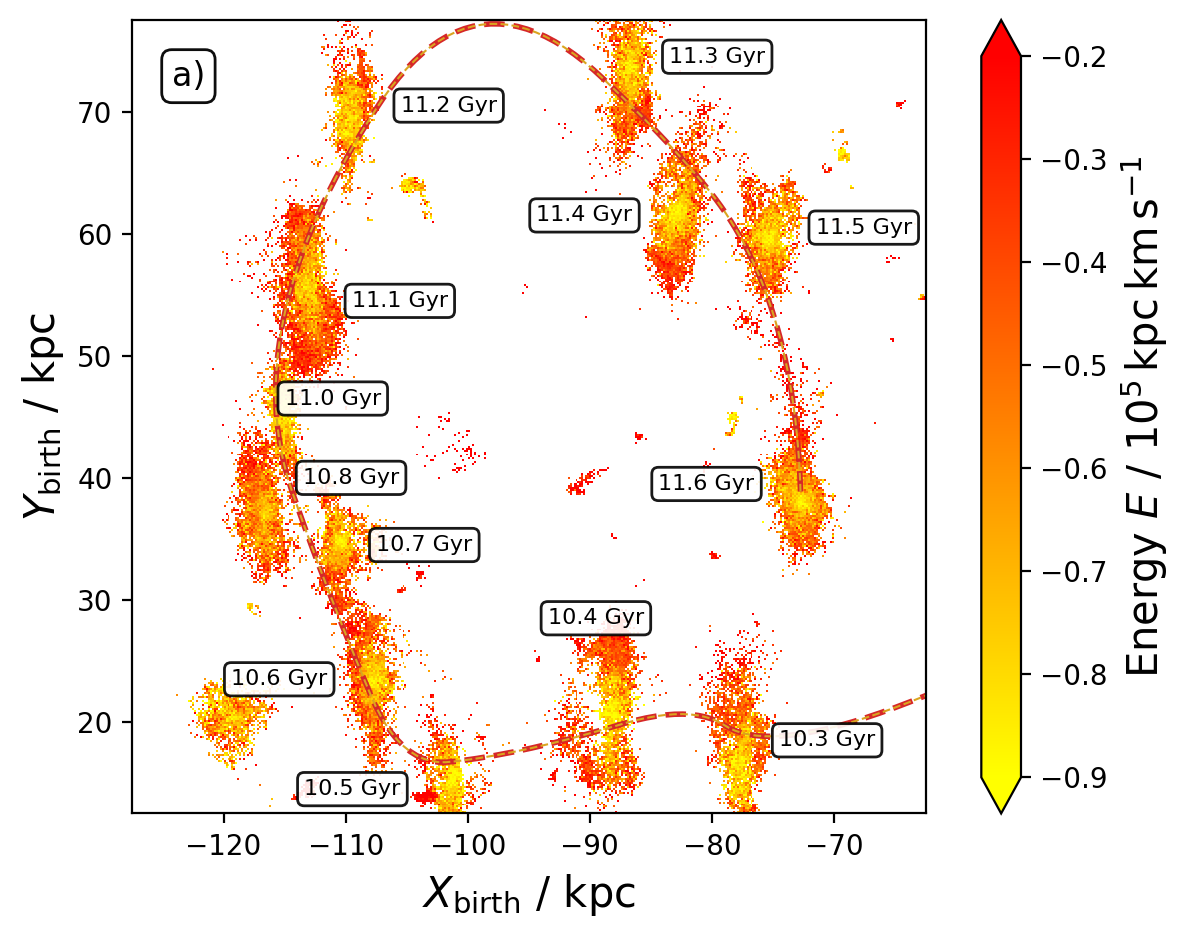}
    \includegraphics[width=\columnwidth]{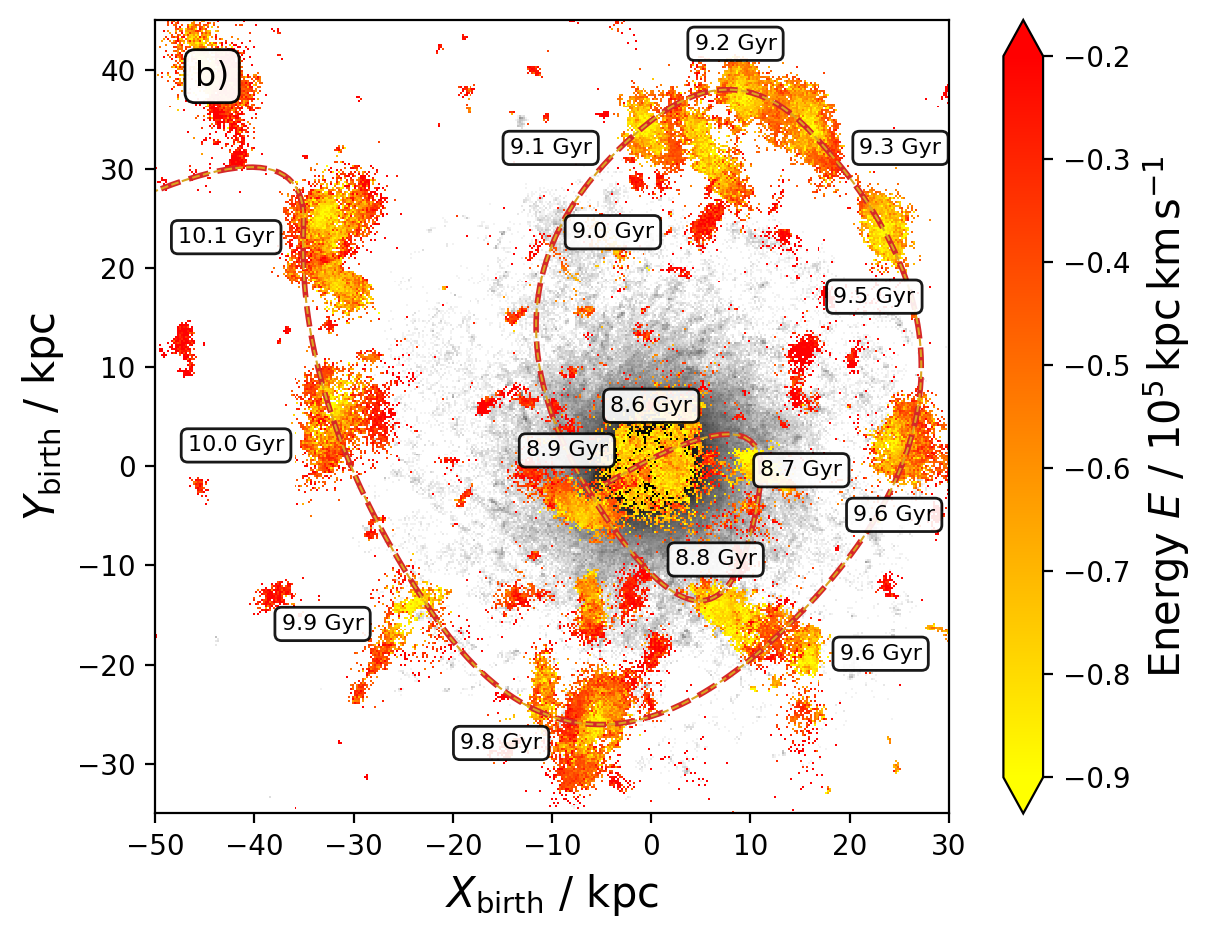}
    \caption{Birth positions (determined in batches of $100\,\mathrm{Myr}$) coloured by present-day orbit\changed{al} energy, both before the major merger (top, for star formation between 11.65 and $10.25\,\mathrm{Gyr}$ ago) and just before and during the merger (bottom, for star formation between 10.15 and $8.55\,\mathrm{Gyr}$ ago), with the main galaxy's birth positions visible in greyscale \href{https://github.com/svenbuder/golden_thread_I/tree/main/figures}{\faGithub}.}
    \label{fig:fellowship_and_mount_doom}
\end{figure}

\subsubsection{Orbit\changed{al} memory of present-day positions}

Having found that the orbit\changed{al} energy is correlated with metallicity and elemental abundances, we are now interested in understanding where the accreted stars with different orbit\changed{al} energies are distributed within the Milky Way analogue. In Figs.~\ref{fig:xy_distribution_ezones}, we therefore show the distribution of the accreted stars from the four energy zones in the $X$ vs. $Y$ projection\footnote{We completeness, we also attach the $X-Z$ projection in Fig.~\ref{fig:xz_distribution_ezones}.}. Since the lower orbit\changed{al} energies correlate with lower radial actions, we of course find that the stars with lower energies are more restricted to the inner galaxy. We find $68\,\mathrm{\%}$ of the stars with $E > -0.37\times10^5\,\mathrm{kpc\,km\,s^{-1}}$ currently have Galactocentric cylindrical radii and heights of $R_\mathrm{2D}$ of $20.2_{-8.0}^{+11.3}\,\mathrm{kpc}$%
 and $z$ of $0.2_{-8.5}^{+8.9}\,\mathrm{kpc}$%
, respectively, whereas the stars with $E < -0.77\times10^5\,\mathrm{kpc\,km\,s^{-1}}$ have radii $R_\mathrm{2D}$ of $3.6_{-1.7}^{+1.7}\,\mathrm{kpc}$%
 and $z$ of $0.0_{-1.3}^{+1.3}\,\mathrm{kpc}$%
, respectively. This means that more metal-rich accreted stars are expected to be found closer to the Galactic centre. We will discuss the implications of this finding for strategies needed to find informative surviving members of the last major merger of the Milky Way in Section~\ref{sec:discussion_strategy_finding_gse_members}.

\subsubsection{Orbit\changed{al} memory of birth positions}

A major motivation of our study is to investigate if there is any memory of the birth positions of accreted stars left in their orbits. \citet{Skuladottir2025} approached this question forward-looking, by tagging the radius of particles within the smaller galaxy in the simulation by \citet{Mori2024} before the merger and then tracing the time of pericenter passages and present-day orbit\changed{al} properties. In a variation of this approach, our post-processing of birth positions allows us to simply use the birth positions and colour them by their present-day orbit\changed{al} energies. By focussing on two specific regions of birth positions in the $X_\mathrm{birth}$ vs. $Y_\mathrm{birth}$ direction in Fig.~\ref{fig:fellowship_and_mount_doom}, we are effectively tracing the star formation (with a resolution of $\sim 100\,\mathrm{Myr}$) of the galaxy long before the merger (Fig.~\ref{fig:fellowship_and_mount_doom}a, showing star formation between 11.65 and $10.25\,\mathrm{Gyr}$ ago), as well as before and during the merger (Fig.~\ref{fig:fellowship_and_mount_doom}b, showing star formation between 10.15 and $8.55\,\mathrm{Gyr}$ ago) with the main galaxy (grey background).

A striking feature of basically all de-facto snapshots of star formation is that the stars born in the outskirts of the progenitor now have high orbit\changed{al} energies (are coloured red), while the stars in the centre of the progenitor in each $100\,\mathrm{Myr}$ snapshot have low orbit\changed{al} energies (yellow). This is fully consistent with the conclusions drawn by \citet{Skuladottir2025} \changed{as well as \citet{Khoperskov2023b}} in the outskirts of the accreted galaxy have higher orbit\changed{al} energies and stars born in its core have lower orbit\changed{al} energies.

\subsubsection{\changed{Quantifying the metallicity gradients with respect to progenitor birth radius and orbital energy}}
\label{sec:birth_radius_gradients}

\begin{figure*}
    \centering
    \includegraphics[width=\linewidth]{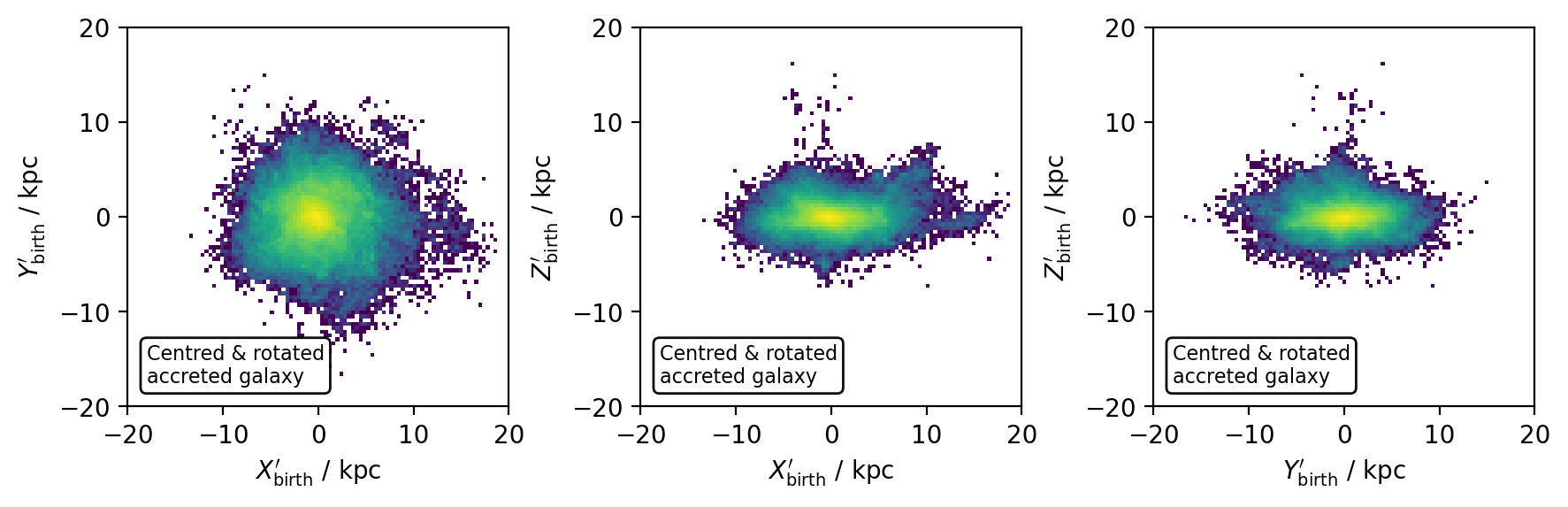}
    \caption{\changed{Birth positions of stars formed in the accreted progenitor galaxy in its centred and rotated reference frame prior to disruption. The three panels show the projected distributions in the $(X_\mathrm{birth}^\prime, Y_\mathrm{birth}^\prime)$, $(X_\mathrm{birth}^\prime, Z_\mathrm{birth}^\prime)$, and $(Y_\mathrm{birth}^\prime, Z_\mathrm{birth}^\prime)$ planes. Colours indicate the logarithmic number density of stellar birth locations} \href{https://github.com/svenbuder/golden_thread_I/tree/main/figures}{\faGithub}.}
    \label{fig:accreted_birth_positions}
\end{figure*}

\changed{To quantify how stellar abundances trace formation location within the accreted progenitor galaxy, we examine the relation between metallicity and projected birth radius $R_\mathrm{birth}^\prime$ as well as between metallicity and present-day orbital energy $E$. To define $R_\mathrm{birth}^\prime$, we express stellar birth positions in a progenitor-centred reference frame and rotate the coordinates using the mass-weighted inertia tensor so that the stellar birth distribution is viewed approximately face-on in the $(x^\prime,y^\prime)$ plane. The resulting spatial distribution of stellar birth positions is shown in Fig.~\ref{fig:accreted_birth_positions}.}

\begin{figure*}
    \centering
    \includegraphics[width=\linewidth]{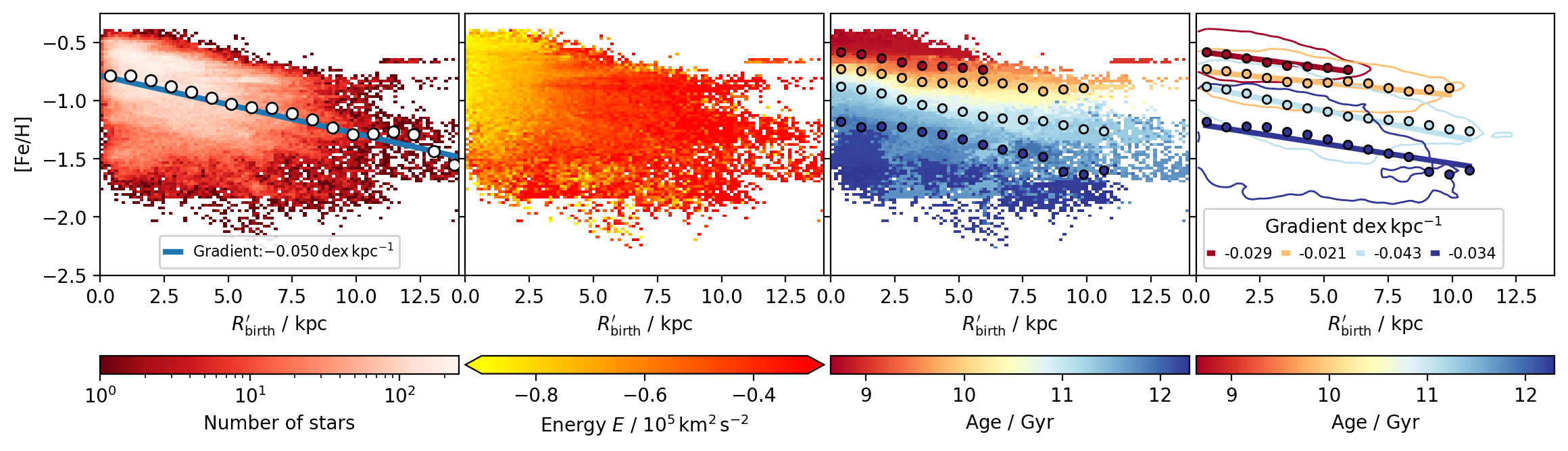}
    \caption{\changed{Relation between projected birth radius $R_\mathrm{birth}^\prime$ and metallicity $\mathrm{[Fe/H]}$ for stars formed in the accreted progenitor. From left to right: stellar number density with median trend and best-fitting radial metallicity gradient; median orbital energy in the $(R_\mathrm{birth}^\prime,\mathrm{[Fe/H]})$ plane; median stellar age with age-binned median sequences; and age-resolved radial metallicity gradients with contours and linear fits for selected age intervals. Colours indicate the logarithmic number density, median energy, or age as labelled} \href{https://github.com/svenbuder/golden_thread_I/tree/main/figures}{\faGithub}.}
    \label{fig:accreted_r_feh_age}
\end{figure*}

\changed{Using this coordinate system, we compute the projected birth radius $R_\mathrm{birth}^\prime = \sqrt{x_\mathrm{birth}^{\prime 2}+y_\mathrm{birth}^{\prime 2}}$ for stars formed in the accreted progenitor prior to disruption. The relation between $R_\mathrm{birth}^\prime$ and metallicity is shown in Fig.~\ref{fig:accreted_r_feh_age}. Considering the full accreted stellar population, we find a clear monotonic decrease of metallicity with increasing birth radius, with a best-fitting gradient of $\frac{\mathrm{d[Fe/H]}}{\mathrm{d}R_\mathrm{birth}^\prime} = -0.05\,\mathrm{dex\,kpc^{-1}}$. This demonstrates that stars forming at larger radii in the progenitor system are systematically more metal-poor, consistent with an ordered chemical structure already present prior to accretion.}

\begin{figure*}
    \centering
    \includegraphics[width=\linewidth]{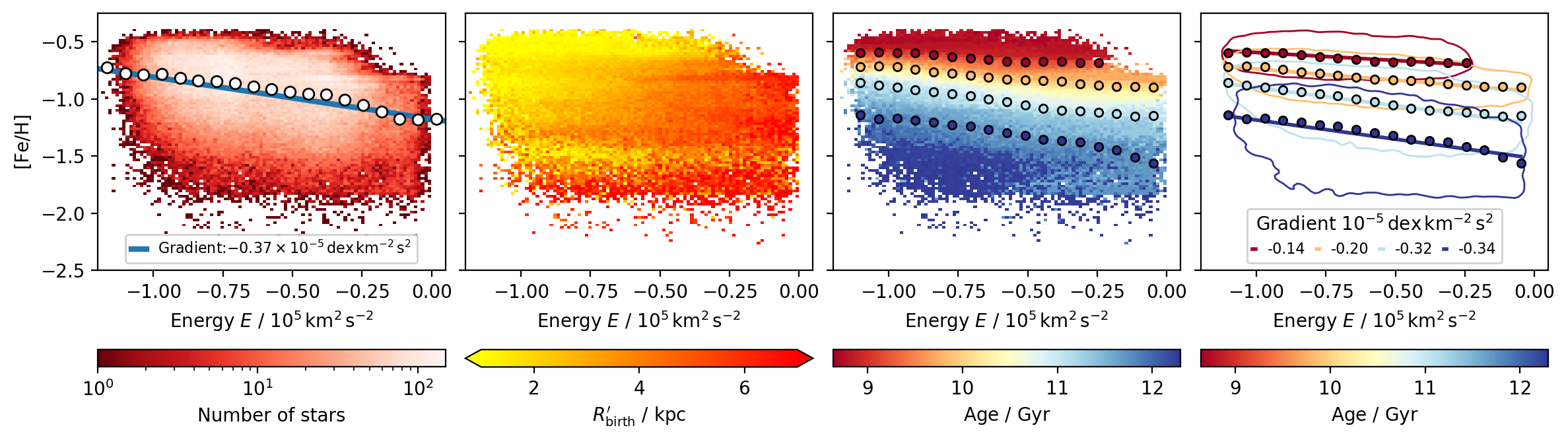}
    \caption{\changed{Relation between orbital energy $E$ and metallicity $\mathrm{[Fe/H]}$ for stars formed in the accreted progenitor. From left to right: stellar number density with median trend and best-fitting metallicity--energy gradient; median projected birth radius $R_\mathrm{birth}^\prime$ in the $(E,\mathrm{[Fe/H]})$ plane; median stellar age with age-binned median sequences; and age-resolved metallicity--energy gradients with contours and linear fits for selected age intervals. Colours indicate the logarithmic number density, median birth radius, or age as labelled} \href{https://github.com/svenbuder/golden_thread_I/tree/main/figures}{\faGithub}.}
    \label{fig:accreted_energy_feh_age}
\end{figure*}

\changed{To isolate the role of chemical evolution within the progenitor, we repeat this analysis in stellar age intervals of width $1\,\mathrm{Gyr}$ between $12\pm0.5$ and $9\pm0.5\,\mathrm{Gyr}$. Within these age-selected subsamples, the metallicity--radius relations become significantly tighter while maintaining similar slopes, with gradients of $-0.034$, $-0.043$, $-0.021$, and $-0.029\,\mathrm{dex\,kpc^{-1}}$ for progressively younger populations. The reduced scatter at fixed age indicates that the broader relation observed for the full population is primarily driven by progressive chemical enrichment over time, while the underlying radial structure of the progenitor galaxy remains similar across stellar generations.}

\changed{We perform an analogous analysis for the relation between metallicity and present-day orbital energy, shown in Fig.~\ref{fig:accreted_energy_feh_age}. For the full accreted population we obtain a metallicity-energy gradient of $\frac{\mathrm{d[Fe/H]}}{\mathrm{d}E} = -0.37\times10^{-5}\,\mathrm{dex\,km^{-2}\,s^{2}}$, with less bound stars being systematically more metal-poor. Age-resolved subsamples again show comparable but slightly shallower gradients of $-0.34$, $-0.32$, $-0.20$, and $-0.14\times10^{-5}\,\mathrm{dex\,km^{-2}\,s^{2}}$, together with a reduction in intrinsic scatter at fixed age.}

\changed{The consistency of these gradients across stellar age intervals indicates that the metallicity structure reflects the internal chemical evolution of the progenitor galaxy rather than being produced by dynamical redistribution after accretion. The measured radial metallicity gradient of
$\mathrm{d[Fe/H]}/\mathrm{d}R_\mathrm{birth}^\prime = -0.05\,\mathrm{dex\,kpc^{-1}}$
is in quantitative agreement with the range of infall gradients reported for \textit{GSE} analogues in the Auriga simulations by \citet{Carrillo2026}, who find values between $-0.14$ and $-0.05\,\mathrm{dex\,kpc^{-1}}$. Similarly, the metallicity-energy gradient measured here,
$\mathrm{d[Fe/H]}/\mathrm{d}E = -0.37\,\mathrm{dex}/10^{-5}\,\mathrm{km^2\,s^{-2}}$,
is consistent with the range $-2.54$ to $-0.42\,\mathrm{dex}/10^{-5}\,\mathrm{km^2\,s^{-2}}$ reported for analogous progenitor systems at infall. Comparable abundance gradients in integrals-of-motion space have also been identified in the HESTIA simulations by \citet{Khoperskov2023c}.}

\section{Discussion}
\label{sec:discussion}

In this paper (Paper~I), we have posed three key questions regarding the information recoverable from integrals-of-motion space. We now address these in turn. In Section~\ref{sec:discussion_finding_accreted_stars}, we discuss the efficiency and biases in selecting accreted stars. In Section~\ref{sec:discussion_strategy_finding_gse_members}, we consider the implications of these findings for future observational strategies to identify informative members of the Milky Way’s last major merger, \textit{GSE}. In Section~\ref{sec:discussion_memory}, we discuss what chemical and birth position memory of the progenitor galaxy is retained in integrals-of-motion space. Together, these results highlight both the opportunities and the limitations in reconstructing the Milky Way’s merger history from present-day phase-space substructures \changed{\citep[see also][]{Thomas2025}}.

\subsection{Selection effects of accreted stars in integrals of motion space} \label{sec:discussion_finding_accreted_stars}

Our investigations in Section~\ref{sec:analysis_dynamic_properties} showed that the true extent of accreted stars in the simulated Milky Way analogue extends far into the inner galaxy, which is dominated by in-situ stars. This dominance explains why current observational research on the extent of accreted stars has identified a rather sharp transition at specific orbit\changed{al} energies or radial actions between in-situ and accreted stars of the \textit{GSE} \citep{Helmi2018, Feuillet2021, Monty2024}.

However, our results show that significantly more accreted stars are present with energies and radial actions below the observationally suggested limits in $E$ \citep{Helmi2018, Monty2024} and $J_R$ \citep{Feuillet2020, Feuillet2021}. Importantly, we find $42\,\mathrm{\%}$%
 and  of the accreted stars in regions that are dominated by in-situ stars on these low-energy orbits. The dynamical overlap of in-situ and accreted structures is significant and spectroscopic, that is, chemical insights are needed to separate in-situ from accreted stars \citep[for example][]{Das2020, Horta2021, Sestito2021, Buder2022}. \changed{A comparable behaviour has been found in the HESTIA cosmological simulations by \citet{Khoperskov2023b}, who showed that accreted stars penetrate deeply into the inner few kiloparsecs and overlap strongly with in-situ populations in $E$–$L_z$ space.}

While several accreted stars and even structures have been found in the inner Galaxy thanks to infrared spectroscopy \citep[for example][]{Horta2021}, \changed{these} regions remains mostly out-of-reach for high-precision optical spectroscopy and partially even astrometry \citep{Queiroz2023}.

\begin{figure}
    \centering
    \includegraphics[width=\columnwidth]{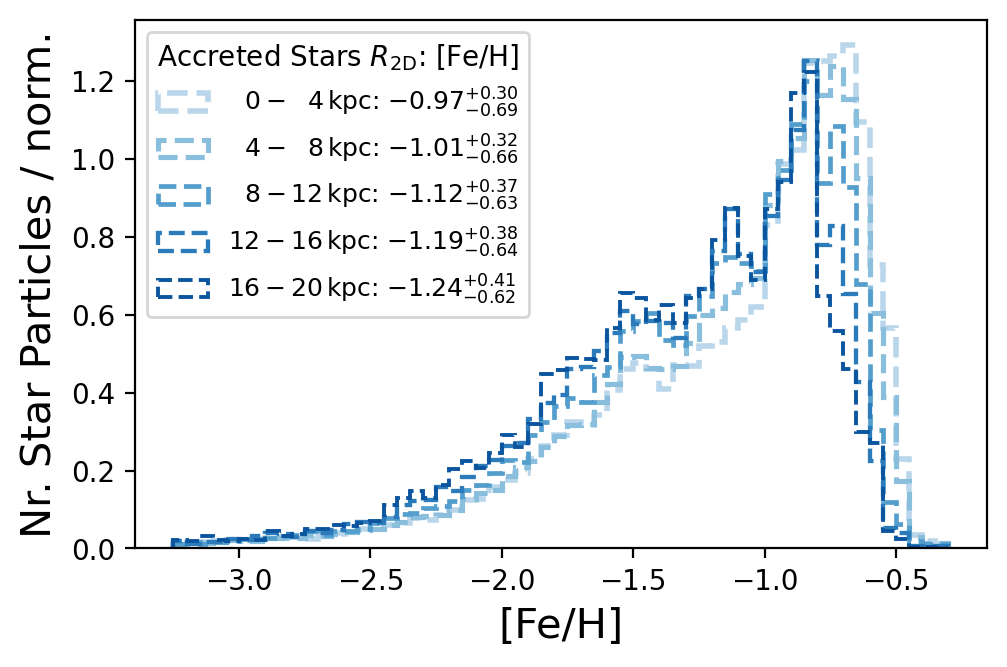}
    \caption{Histograms of {[Fe/H]} distribution for accreted stars with different present-day galactocentric radii $R_\mathrm{2D}$ with median {[Fe/H]} indicated in the legend for a each region \href{https://github.com/svenbuder/golden_thread_I/tree/main/figures}{\faGithub}.}
    \label{fig:fe_h_histograms_r_bins}
\end{figure}

The current dedicated spectroscopic searches for accreted stars, which are typically limited to either the nearby or extended Solar neighbourhood \citep[for example][]{Nissen2010, Buder2022} or far away from the Galactic disc \citep{Naidu2020} can provide significant insights, but are expected to be not only incomplete but also significantly biased because of the subsample of \textit{GSE} stars that they are able to target. We remind ourselves that in the simulation, the spatial distribution of the lowest-energy accreted stars is limited to within $R_\mathrm{2D} = $ and $Z = $ (see Figs.~\ref{fig:xy_distribution_ezones} and \ref{fig:xz_distribution_ezones}). Assuming the distribution of accreted stars would be the same for the Milky Way, we would thus not expect a lot (or any) of the accreted stars with lowest energies to be present in the sample of 33 accreted stars in the Solar neighbourhood by \citet{Nissen2010}, but mainly those with $E > -0.57\times10^5\,\mathrm{kpc\,km\,s^{-1}}$, consistent with the sample studied by \citet{Skuladottir2025}. The same holds true for Milky Way halo studies \citep{Naidu2020}, which have helped us to better understand the substructure of accretion, but are expected to mostly target \textit{GSE} stars with $E > -0.77\times10^{5}\,\mathrm{kpc\,km\,s^{-1}}$. More generally, we expect different [Fe/H] distributions to be recorded depending on the heights and radii of targeted accreted stars, as shown for a selection of radii in Fig.~\ref{fig:fe_h_histograms_r_bins}. This could potentially contribute to the differences of the [Fe/H] distribution of the observed \textit{GSE} by \citet{Das2020}, \citet{Naidu2020}, \citet{Feuillet2020, Feuillet2021}, and \cite{Buder2022} who analysed the different observations of the APOGEE, H3, SkyMapper, \textit{Gaia}, and GALAH surveys \citep[for a comparison of their reported {[Fe/H] distributions see Fig.~10 by}][]{Buder2022}.

Our simulation data suggests that we will find a currently under-represented subset of stars in the inner Galaxy, more specifically within its bulge region. Over the decades \citep[see for example][for a review]{Barbuy2018}, this dense region has been the target of several dedicated spectroscopic studies \citep[for example][]{Ness2013, Bensby2017, Lucey2019}. 
This raises the question \changed{whether some of these accreted stars may already have been identified observationally.} \citet{Ness2013, Ness2013b}, for example, have identified specific overdensities in the metallicity distribution functions of the bulge region (i.e.\ below the centre at $b = -5\,\mathrm{deg}$), namely at $\mathrm{[Fe/H]} \in [-1.73, -1.16, -0.66, -0.26, +0.12]$ with typical spreads of $\pm0.11$–$0.15$ \citep[see also][]{Portail2017}.
Despite some differences between the actual peak positions \citep[compare for example Fig.~4 by][]{Barbuy2018}, different studies agreed with the presence of multiple overdensities. \citet{Bensby2017} found the second-lowest metallicity peak around $\mathrm{[Fe/H]} = -1.09$. \citet{Portail2017} also found a peak at $-1.18\pm 0.14$ with less (to no) net rotation, reminiscent of the characteristic [Fe/H] and $L_Z$ distribution of accreted stars. 

\begin{figure}
    \centering
    \includegraphics[width=\columnwidth]{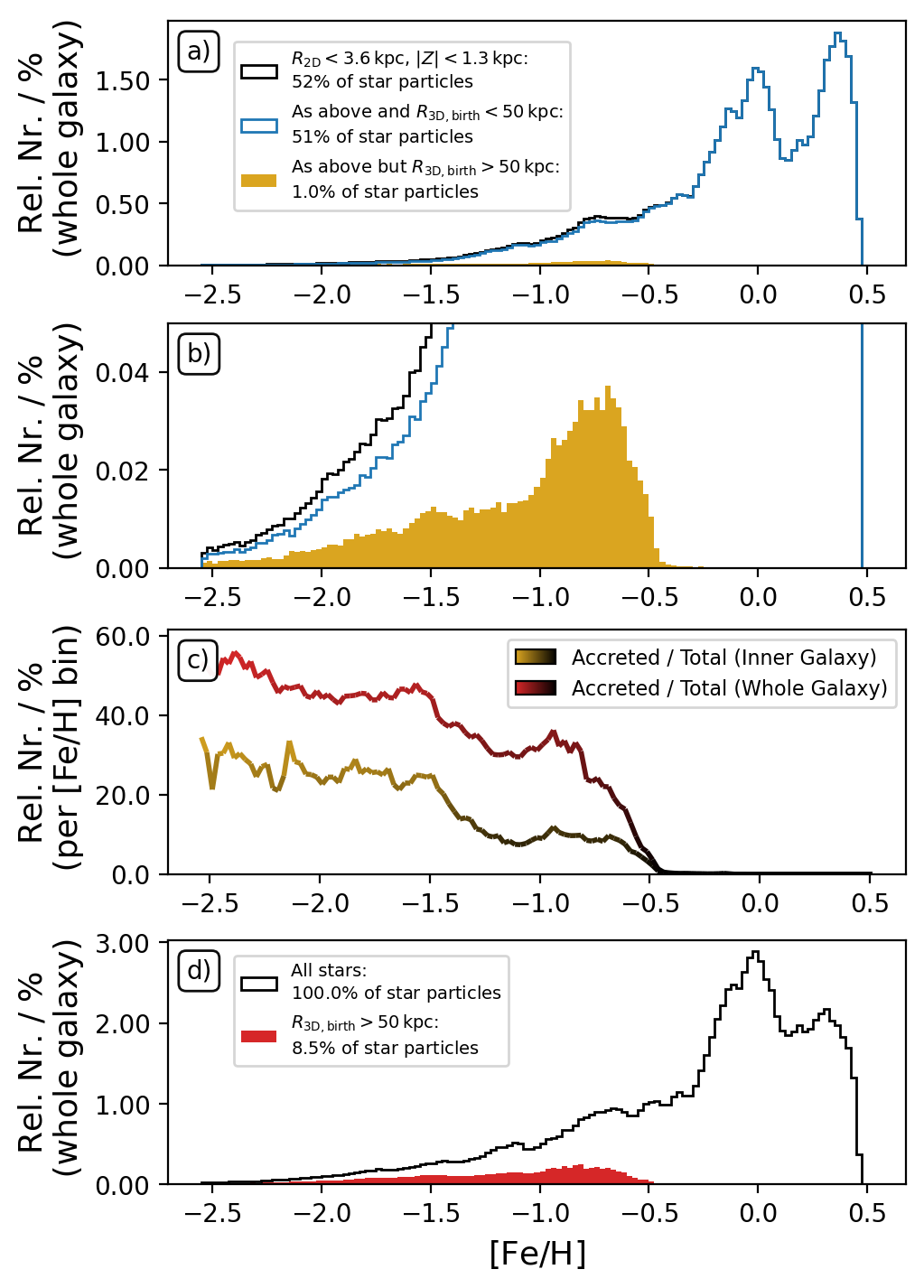}
    \caption{Metallicity distribution functions (relative to total number of stars in the galaxy). Top panel shows the distribution of the inner $R_\mathrm{3D} < 3.6\,\mathrm{kpc}$ and how in-situ (blue) and accreted populations (golden) contribute to it. Panel b) is showing the same golden distribution of accreted stars as panel a), but zooming vertically in to see more details. Panel c) traces the relative fraction of accreted stars to all stars in the inner galaxy (yellow-black line) as well as the galaxy overall (red-black line). Panel d) shows the distribution of the whole galaxy (black) and all accreted stars (red) \href{https://github.com/svenbuder/golden_thread_I/tree/main/figures}{\faGithub}.}
    \label{fig:fe_h_histogram_inner_galaxy}
\end{figure}

When focusing on the inner region of the Milky Way analogue ($R_\mathrm{2D} < 3.6\,\mathrm{kpc}$ and $\vert Z\vert < 1.3\,\mathrm{kpc}$), we also find a multi-modal metallicity distribution (see Fig.~\ref{fig:fe_h_histogram_inner_galaxy}a). The distribution is overall shifted to higher [Fe/H] values than the Milky Way by $\sim+0.3\,\mathrm{dex}$ compared to the measurements in the Milky Way bulge (see above). Taking this into account, we would expect the accreted stars to contribute mainly to the peak around $\sim-0.8$. We can confirm this when selecting the accreted stars of the simulation (golden in Fig.~\ref{fig:fe_h_histogram_inner_galaxy}a and zoomed in vertically for better visibility in Fig.~\ref{fig:fe_h_histogram_inner_galaxy}b). What strikes us here is the limited contribution of accreted stars to the inner galaxy (only $2\,\mathrm{\%}$%
) but especially to the peak itself. This suggests two things: firstly that the \changed{number} of accreted stars distributed in the inner galaxy is insignificant (on a relative level) and secondly that there are still stars formed with the same, outstanding iron abundance as the accreted stars. A simple explanation could be that these particular stars formed from gas of the accreted galaxy -- an exciting avenue for quantifying the amount of metal-poor gas brought in by the accreted galaxy \citep{Buck2023}. This hypothesis should be followed up in a future study that dives into the detailed tracing of the gas of the simulation during the merger. In any case, we do note that the number of stars born with this iron abundance -- and in particular those contributing to the bump around $\mathrm{[Fe/H] \sim -0.75}$ in the simulation -- appears insignificant compared to the rest of stars in this inner region. A blind survey of 100 stars in this region would be expected to only find 2 accreted stars on average, whereas a survey that is able to preselect stars of the inner galaxy with $\mathrm{[Fe/H]} < -0.5$ would be able to increase the sampling rate by at least a factor of 5 to $10-30\,\%$ (Fig.~\ref{fig:fe_h_histogram_inner_galaxy}c). Although the overall contribution of accreted stars to the present-day stellar distribution is around 9\% (see Fig.~\ref{fig:fe_h_histogram_inner_galaxy}d), blindly sampling accreted stars is more successful in regions with less metal-rich stars, such as the Galactic halo, as already well established by observations of the Milky Way \citep[e.g.][]{Conroy2019}.

The question therefore arises, if we perhaps have not actually \changed{observed a significant fraction of} the core of the \textit{GSE} progenitor yet, since \changed{it is} so well hidden in the centre of our Galaxy. While focussing on a larger metallicity-range in the metal-poor inner bulge, we note that \citet{Lucey2022} only found a small number and percentage of stars with accretion-like, sub-solar [Al/Fe] and [Mn/Fe] in their sample (see their Figs.~7 and 8). Such stars certainly should be followed up to identify if they are part of the accreted population of the \textit{GSE} \citep[see also][]{Kunder2025}.

An observation that has previously been debated as a potential finding of the core of the last major merger of the Milky Way was the Heracles substructure \citep{Horta2021}, with some of the lowest energies of accreted stars found so far. While some researchers have argued that its seemingly distinct separation from the \textit{GSE} in the $L_Z$ vs. $E$ could have been caused by bar interactions \citep{Dillamore2025}, our findings from the simulation suggest that its metallicity (relative to the \textit{GSE}) is in fact too low to be part of the same system. We thus agree with \citet{Horta2021} that this system is likely a separate accretion event within the Milky Way.

Looking into the near future, the 4MOST survey's 4MIDABLE consortium surveys of the inner galaxy \citep{4MOST_HR_DiskBulge, 4MOST_LR_DiskBulge} will observe millions of stars towards this region. If the last major merger in our Milky Way is comparable to the one in our analogue simulation, we expect these surveys to observe a considerable number of accreted stars. More generally looking at observations of the inner galaxy, if one could preselect stars with $\mathrm{[Fe/H]} < -0.5$ for spectroscopic follow-up to measure Na or Al as well as Mg and Mn, our simulation data suggests an increase of the ratio of accreted stars in the inner Galaxy from  with a blind selection to $10\,\mathrm{\%}$%
 with a metallicity-cut. Preselection methods based on photometric data (e.g. Pristine, SkyMapper; \citealt{Starkenburg2017, DaCosta2019}) or on spectrophotometric combinations of \textit{Gaia} BP/RP spectra and Pristine photometry \citep{Martin2024} have already proven effective in the inner Galaxy \citep[e.g. with the Pristine Inner Galaxy Survey;][]{Arentsen2020, Arentsen2020b}. Comparable strategies could be applied to future infrared surveys such as APOGEE/SDSS \citep{Santana2021}. Subsequently we discuss why such observations could be key for our understanding of the last major merger of the Milky Way by allowing us to better quantify the amount of low-energy \textit{GSE} members.

\subsection{The importance of finding low-energy \textit{GSE} members} \label{sec:discussion_strategy_finding_gse_members}

What could be the potential impact of missing out on the hidden low-energy members of the \textit{GSE}? The detailed impact of the major merger will lie in the impact of the accreted gas and stars onto the chemical make-up and rotation of the Galaxy as well as the merger configuration and pathway. Two general galaxy properties -- galaxy mass or mean metallicity -- are often used to quantify a galaxy and its potential impact during a merger. With the observationally found mass-metallicity relation \citep[for example][]{Gallazzi2005, Kirby2013}, these numbers are correlated and allow to roughly estimate the mass of a system from its metallicity and vice versa. In the Milky Way, this relation has been confirmed with surviving smaller galaxies \citep{Kirby2013} and then used to estimate the masses of disrupted galaxies by comparison \citep[for example][]{Helmi2018} -- in addition to dynamical mass estimates from simulations that recovered the dynamical signatures of the disrupted galaxies \citep[for example][]{Naidu2022b}.

\begin{figure}
    \centering
    \includegraphics[width=\columnwidth]{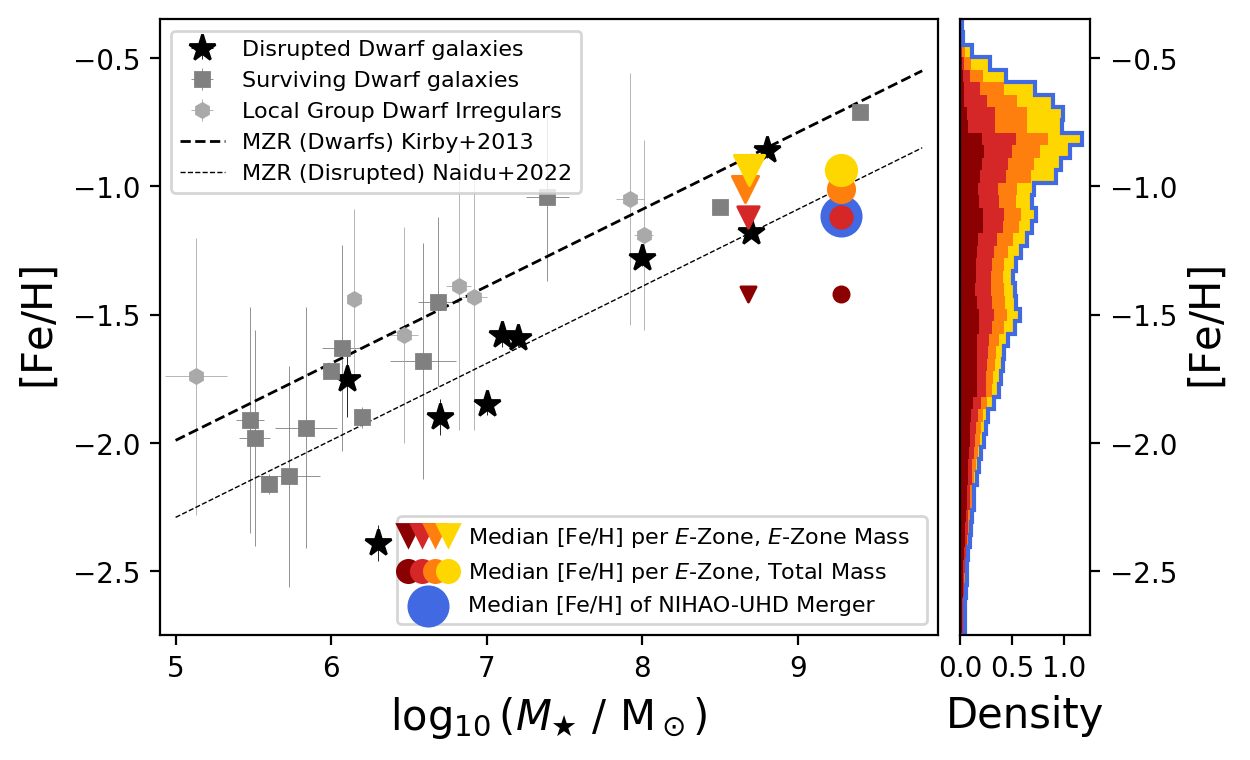}
    \caption{Mass-metallicity relations (dashed lines) for different galaxies and selections. We list both reported measurements of disrupted (black star symbols), surviving  (grey squares), and irregular dwarf galaxies (grey circles) by \citet{Kirby2013} and \citet{Naidu2022b}. For the major merger of the NIHAO-UHD Milky Way analogue we show both the median [Fe/H] and total mass for the whole merger (blue dot), and median [Fe/H] of the different energy zones vs. the total merger mass (coloured dots); and only the masses of star particles in the individual energy zones (coloured triangles). Right panel shows the stacked histograms of the different metallicity distributions following the same colour code \href{https://github.com/svenbuder/golden_thread_I/tree/main/figures}{\faGithub}.}
    \label{fig:mzr_different_selections}
\end{figure}

By comparing the mass-metallicity relations of observed surviving and (partially or fully) disrupted dwarf galaxies, \citet{Naidu2022b} uncovered a significant offset between them, with the latter's metallicities being on average 0.30\,dex lower. This difference challenges the assumption of a universal mass–metallicity relation. However, if the preselection methods discussed above preferentially include or exclude stars of certain metallicities, part of the observed offset could arise from selection effects, in addition to potential differences stemming from continued chemical enrichment in surviving systems compared to quenched evolution in disrupted ones, \changed{as seen for example in the HESTIA simulations by \citet{Khoperskov2023c}}. Beyond the biases for \textit{GSE} \citep{Skuladottir2025}, we are wondering if the [Fe/H] estimates of surviving dwarf galaxies could be biased by the analysis of stars in their cores (and less so their outskirts), whereas the [Fe/H] estimates of disrupted dwarf galaxies could be biased towards lower values by the stars in their previous outskirts, rather than their more metal-rich cores (which our simulations suggest to be more likely embedded in the inner galaxy regions dominated by in-situ stars). To visualise this point, we are recreating the mass-metallicity relation as shown by \citet{Naidu2022b} in Fig.~\ref{fig:mzr_different_selections}. To this figure, we add the median [Fe/H] values for the four different energy zones of the accreted stars in our simulation (with yellow to dark red corresponding to the lowest to highest energy zones from Fig.~\ref{fig:fe_h_histograms}). We indeed see that when selecting preferentially (or even only) the higher energy samples, we tend to infer a lower [Fe/H]. The lower-energy samples would increase the average median [Fe/H]. We also notice that shape -- in particular the non-Gaussianity -- of the metallicity distribution functions of a galaxy needs to be included into this figure for both surviving and disrupted dwarf galaxies, rather than the currently used mean or median [Fe/H].

To estimate the potential impact of systematic offsets of metallicities, we invert the published stellar mass–metallicity relations for dwarf galaxies. Following \citet{Naidu2022b}, we adopt either their relation for disrupted dwarfs (their Eq.~3),
\begin{align}
\qquad\qquad    \log_{10} (M_\star/\mathrm{M_\odot}) &= 2.78 \times \mathrm{[Fe/H]} + 11.86 , \label{eq:naidu}
\end{align}
or the relation for surviving dwarfs from \citet{Kirby2013}, shifted by $-0.30\,\mathrm{dex}$,
\begin{align}
\qquad\qquad    \log_{10} (M_\star/\mathrm{M_\odot}) &= 3.33 \times \mathrm{[Fe/H]} + 12.63 . \label{eq:kirby}
\end{align}

Inserting the median $\mathrm{[Fe/H]}$ values of the four orbital-energy zones, we infer the logarithmic stellar masses of $\log_{10}(M_\star/\mathrm{M_\odot}) \simeq 7.9$–$9.5$ listed in Tab.~\ref{tab:energy_selection}. This range spans $\sim$1.6 dex, that is, a factor of $\sim$40 in stellar mass, depending on which part of the progenitor is sampled. 

This exercise illustrates that chemically inferred stellar masses are highly sensitive to the adopted metallicity distribution and to the choice of mass–metallicity relation. In a future project, we will therefore re-assess the mass–metallicity relation in detail, accounting for full metallicity distributions and for potential systematic differences between disrupted and surviving dwarf galaxies.

\subsection{Chemical and birth position memory of the progenitor galaxy in integrals of motion space} \label{sec:discussion_memory}

Throughout our analysis in Section~\ref{sec:analysis_chemodynamic_memory}, we have investigated if we can find the same patterns from the observational study by \citet{Skuladottir2025} that accreted stars with different orbit\changed{al} energies carry both chemical and birth position memory.

Indeed, we have found correlations of present-day orbit\changed{al} energy and metallicity (Fig.~\ref{fig:fe_h_histograms}) as well as chemical enhancement of several individual elements (Fig.~\ref{fig:xfe_feh_zones}), but also a correlation with present-day orbits (Figs.~\ref{fig:xy_distribution_ezones} and \ref{fig:xz_distribution_ezones}) and most importantly birth positions (Fig.~\ref{fig:fellowship_and_mount_doom}). While these findings are all in agreement with the findings by \citet{Skuladottir2025} at face value, here we want to discuss a few more subtle differences that we note because of our larger sample size that also reaches towards the lower-energy regions and thus the core of the major merger galaxy.

Because of the low number of only 33 targets, \citet{Skuladottir2025} could only subdivide their sample into two groups. To quantify the abundance differences between the groups beyond computing abundance offsets, they fitted linear functions in [Fe/H] vs. [X/Fe] space and were able to identify both overall offsets and linear, that is, metallicity-dependent trends. However, the abundance trends of our simulation in Fig.~\ref{fig:xfe_feh_zones} often show non-linear trends, which would be insufficiently covered by linear functions. In particular, we note a shift of the skewed metallicity distribution function with decreasing orbit\changed{al} energy from metal-poor to metal-rich. \citet{Skuladottir2025} have in fact made a similar discovery in their research, when assessing the impact of the energy cut onto the metallicity distribution function (see their Appendix~A), where moving their energy cut to lower energies limits the sample to quite high $\mathrm{[Fe/H]} \gtrsim -0.9$ and reveals two significantly different distributions, in line with our findings. Again in line with our findings, they found significantly different abundance patterns, especially for $\mathrm{[Fe/H]} > -1$.

Because of the changing metallicity distribution function and some non-linear trends in [X/Fe] in our larger -- but simulated -- sample, we believe that future investigation of observed abundance trends -- in the best case with larger samples across a diverse energy range -- should aim to go beyond the fitting of linear trends to these distributions as done by \citet{Skuladottir2025}.

\subsection{\changed{Metallicity gradient of the \textit{GSE} progenitor}}
\label{sec:discussion_gradient}

Our investigations confirm the conclusions drawn by \citet{Skuladottir2025}, that the lower-energy stars were born in the inner parts of the now-accreted galaxy, whereas higher-energy stars were born in its outskirts. Taken together, \changes{the relation between progenitor birth radius, present-day orbital energy, and chemistry suggests that the \textit{GSE} progenitor was a disc galaxy with a measurable internal radial metallicity gradient prior to its merger with the Milky Way.}

\changes{We directly measure this metallicity gradient with respect to progenitor birth radius as $\mathrm{d[Fe/H]}/\mathrm{d}R_\mathrm{birth}^\prime = -0.05\,\mathrm{dex\,kpc^{-1}}$,} consistent with gradients inferred for \textit{GSE} analogues in the Auriga simulations by \citet{Carrillo2026}, who report values between $-0.14$ and $-0.05\,\mathrm{dex\,kpc^{-1}}$, and with the structured metallicity variations across integrals-of-motion space identified in the HESTIA simulations by \citet{Khoperskov2023c}.

Our \changes{cosmological} zoom-in simulation thus quantitatively reinforces these results and indicates that the preservation of metallicity gradients with respect to formation location is a robust feature across independent simulation suites. This provides rare direct evidence of internal chemical structure in an accreted galaxy and further strengthens the picture that the \textit{GSE} progenitor likely experienced extended, inside-out star formation before merging with the Milky Way.

\section{Conclusions}
\label{sec:conclusions}

We have used a high-resolution NIHAO-UHD cosmological simulation to study the last major merger of a Milky Way analogue, addressing three key questions:

First, the present-day orbital properties of accreted stars correlate with their birth radii within the progenitor galaxy (Sec.~\ref{sec:discussion_finding_accreted_stars}), \changed{consistent with similar findings from the HESTIA simulations \citep{Khoperskov2023b}. In particular, stars originating from the chemically enriched core of the progenitor occupy regions of integrals-of-motion space that are typically excluded by standard selection criteria, demonstrating that commonly used dynamical selections recover only a subset of the accreted population.}

Second, these correlations arise from abundance gradients within the progenitor\changed{, which we quantify as
$\mathrm{d[Fe/H]}/\mathrm{d}R_\mathrm{birth}^\prime = -0.05\,\mathrm{dex\,kpc^{-1}}$,
in agreement with the range of gradients measured for \textit{GSE} analogues in the Auriga simulations by \citet{Carrillo2026} and consistent with structured abundance variations across integrals-of-motion space identified in the HESTIA simulations \citet{Khoperskov2023d, Khoperskov2023c}. This demonstrates that stellar abundances can be used to infer formation location within the disrupted progenitor galaxy.} 
Lower-energy stars, now more strongly bound to the Milky Way, formed in the core of the accreted galaxy and are systematically more metal-rich and chemically enhanced. This supports the scenario of \citet{Skuladottir2025} and demonstrates that chemodynamical memory is preserved across the merger. However, accreted stars also extend to much lower orbital energies and radial actions than previously inferred. As a consequence, selection methods in integrals-of-motion space \citep{Helmi2018, Feuillet2021, Monty2024} systematically miss stars from the progenitor’s core. Progenitor reconstructions are therefore biased toward the chemically metal-poor outskirts. These biases may even help explain differences in the mass–metallicity relation between surviving and disrupted dwarf galaxies \citep{Naidu2022}, offering a testable prediction for future work (Secs.~\ref{sec:discussion_memory} and \ref{sec:discussion_strategy_finding_gse_members}).

Third, our analysis shows that current selection functions alone are insufficient to capture the full abundance distribution of the progenitor. Future survey strategies must target inner-Galaxy stars to recover the missing, chemically enriched core population.

Finally, our findings highlight the power of cosmological simulations to quantify biases introduced by observational selection functions, and \changed{demonstrate that spatial chemical abundance gradients set at formation are preserved consistently among simulations. These results subsequently guide} strategies for building a more representative census of accreted stars in the Milky Way.

\section*{Acknowledgments}

We acknowledge the traditional owners of the land on which the ANU stands, the Ngunnawal and Ngambri people. We pay our respects to elders past, and present and are proud to continue their tradition of surveying the night sky and its mysteries to better understand our Universe. SB acknowledges support from the Australian Research Council under grant number DE240100150.
TB's contribution to this project was made possible by funding from the Carl Zeiss Stiftung. \'{A}S~acknowledges funding from the European Research Council (ERC) under the European Union’s Horizon 2020 research and innovation programme (grant agreement No. 101117455).

\section*{Data Availability}

All code to reproduce the analysis and figures can be publicly accessed via \url{https://github.com/svenbuder/golden_thread_I}.
The used simulation snapshot can be publicly accessed as FITS file via \url{https://github.com/svenbuder/preparing_NIHAO}. Original data, more snapshots and other galaxies can be found at \url{https://tobias-buck.de/#sim_data}. We encourage interested readers to get in contact with the authors for full data access and advice for use and cite \citet{Buck2020b, Buck2021}.

\section*{Software}

\textsc{python} (version 3.12.11) and included its packages
\textsc{astropy} \citep[v. 7.1.0;][]{Robitaille2013,PriceWhelan2018},
\textsc{hyppo} \citep[v. 0.5.2;][]{hyppo},
\textsc{IPython} \citep[v. 9.1.0;][]{ipython},
\textsc{matplotlib} \citep[v. 3.10.3;][]{matplotlib},
\textsc{NumPy} \citep[v. 2.2.6;][]{numpy},
\textsc{scipy} \citep[v. 1.16.0;][]{Scipy};
\textsc{topcat} \citep[v. 4.7;][]{Taylor2005};
 
\bibliographystyle{mnras}
\bibliography{bib}

\appendix

\section{Additional Information}

\begin{table*}
    \centering
    \caption{Spatial and age selection used to identify individual (and clean) overdensities in birth positions for Fig.~\ref{fig:tracing_xyz_birth_3} \href{https://github.com/svenbuder/golden_thread_I/tree/main/figures}{\faGithub}.}
    \renewcommand{\arraystretch}{0.9}
    \begin{tabular}{cccccccccc}
\hline \hline
Property & $X$ Range & $Y$ Range & $Z$ Range & Age Range & $X$ Center & $Y$ Center & $Z$ Center & Median Age & Median [Fe/H] \\
Unit & $\mathrm{kpc}$ & $\mathrm{kpc}$ & $\mathrm{kpc}$ & $\mathrm{Gyr}$ & $\mathrm{kpc}$ & $\mathrm{kpc}$ & $\mathrm{kpc}$ & $\mathrm{Gyr}$ & $\mathrm{dex}$ \\
\hline
1 & -40..-25 & 10..33 & -175..-155 & 11.0..13.0 & -32 & 20 & -166 & $12.24\pm0.03$ & $-1.50\pm0.18$ \\
2 & -19..-9 & 46..68 & -174..-160 & 10.9..12.9 & -13 & 56 & -166 & $11.92\pm0.03$ & $-1.27\pm0.12$ \\
3 & 0..12 & 20..45 & -176..-164 & 10.8..12.8 & 6 & 32 & -169 & $11.79\pm0.03$ & $-1.21\pm0.17$ \\
4 & -53..-38 & 30..60 & -166..-144 & 10.7..12.7 & -43 & 39 & -160 & $11.72\pm0.04$ & $-1.36\pm0.24$ \\
5 & -80..-65 & 32..52 & -154..-130 & 10.6..12.6 & -73 & 39 & -143 & $11.60\pm0.04$ & $-1.17\pm0.15$ \\
6 & -79..-72 & 50..66 & -142..-124 & 10.5..12.5 & -75 & 60 & -132 & $11.48\pm0.04$ & $-1.12\pm0.16$ \\
7 & -86..-79 & 50..75 & -132..-114 & 10.4..12.4 & -83 & 61 & -121 & $11.37\pm0.04$ & $-1.14\pm0.17$ \\
8 & -92..-80 & 65..90 & -115..-98 & 10.3..12.3 & -87 & 74 & -106 & $11.29\pm0.03$ & $-1.12\pm0.18$ \\
9 & -114..-106 & 64..81 & -86..-72 & 10.2..12.2 & -109 & 70 & -80 & $11.16\pm0.03$ & $-0.99\pm0.20$ \\
10 & -120..-107 & 45..64 & -86..-66 & 10.1..12.1 & -113 & 54 & -75 & $11.06\pm0.04$ & $-1.04\pm0.15$ \\
11 & -117..-112 & 39..51 & -71..-62 & 9.9..11.9 & -115 & 46 & -67 & $10.96\pm0.03$ & $-0.92\pm0.14$ \\
12 & -120..-114 & 29..46 & -62..-44 & 9.8..11.8 & -117 & 38 & -52 & $10.85\pm0.02$ & $-0.99\pm0.12$ \\
13 & -114..-100 & 29..46 & -62..-44 & 9.7..11.7 & -111 & 34 & -52 & $10.72\pm0.03$ & $-0.96\pm0.12$ \\
14 & -113..-103 & 14..30 & -61..-43 & 9.6..11.6 & -108 & 23 & -50 & $10.62\pm0.04$ & $-0.98\pm0.12$ \\
15 & -104..-98 & 8..25 & -50..-43 & 9.5..11.5 & -102 & 14 & -49 & $10.51\pm0.04$ & $-0.92\pm0.10$ \\
16 & -93..-84 & 12..32 & -66..-48 & 9.4..11.4 & -88 & 23 & -57 & $10.42\pm0.03$ & $-0.94\pm0.10$ \\
17 & -82..-73 & 11..29 & -63..-49 & 9.3..11.3 & -78 & 18 & -58 & $10.31\pm0.03$ & $-0.90\pm0.09$ \\
18 & -62..-53 & 9..34 & -69..-54 & 9.2..11.2 & -57 & 25 & -62 & $10.19\pm0.04$ & $-0.83\pm0.08$ \\
19 & -36..-20 & 17..33 & -73..-44 & 9.1..11.1 & -33 & 25 & -63 & $10.09\pm0.04$ & $-0.79\pm0.10$ \\
20 & -36..-20 & -10..15 & -73..-44 & 9.0..11.0 & -32 & 4 & -54 & $9.98\pm0.04$ & $-0.86\pm0.06$ \\
21 & -31..-19 & -25..-5 & -43..-28 & 8.9..10.9 & -24 & -14 & -36 & $9.88\pm0.04$ & $-0.74\pm0.09$ \\
22 & -12..0 & -35..-15 & -23..-6 & 8.8..10.8 & -6 & -26 & -15 & $9.75\pm0.02$ & $-0.81\pm0.07$ \\
23 & 9..19 & -34..-12 & 2..16 & 8.6..10.6 & 14 & -17 & 7 & $9.61\pm0.43$ & $-0.66\pm0.12$ \\
24 & 20..35 & -6..10 & 14..34 & 8.5..10.5 & 26 & 2 & 23 & $9.56\pm0.02$ & $-0.77\pm0.06$ \\
25 & 14..28 & 19..28 & 25..36 & 8.4..10.4 & 24 & 24 & 30 & $9.46\pm0.03$ & $-0.72\pm0.05$ \\
26 & 10..20 & 27..41 & 23..38 & 8.3..10.3 & 16 & 34 & 29 & $9.34\pm0.03$ & $-0.70\pm0.05$ \\
27 & 6..15 & 31..41 & 15..27 & 8.2..10.2 & 10 & 37 & 22 & $9.22\pm0.02$ & $-0.68\pm0.05$ \\
28 & -4..15 & 28..39 & 8..17 & 8.1..10.1 & -1 & 34 & 14 & $9.10\pm0.04$ & $-0.64\pm0.05$ \\
29 & -17..3 & 12..29 & -8..7 & 8.0..10.0 & -8 & 23 & -1 & $9.04\pm0.03$ & $-0.65\pm0.04$ \\
30 & -19..5 & -13..1 & -12..0 & 7.9..9.9 & -7 & -4 & -6 & $8.89\pm0.04$ & $-0.60\pm0.08$ \\
31 & 4..14 & -21..-10 & 3..10 & 7.8..9.8 & 8 & -15 & 8 & $8.80\pm0.03$ & $-0.56\pm0.04$ \\
32 & 3..27 & -6..3 & 2..11 & 7.7..9.7 & 11 & -1 & 5 & $8.70\pm0.03$ & $-0.50\pm0.03$ \\
33 & -23..19 & -16..13 & -4..5 & 7.6..9.6 & -5 & -2 & 0 & $8.62\pm0.36$ & $-0.49\pm0.07$ \\
\hline \hline
\end{tabular}

    \label{tab:birth_position_tabular}
\end{table*}

\begin{figure*}
    \centering
    \includegraphics[width=0.47\linewidth]{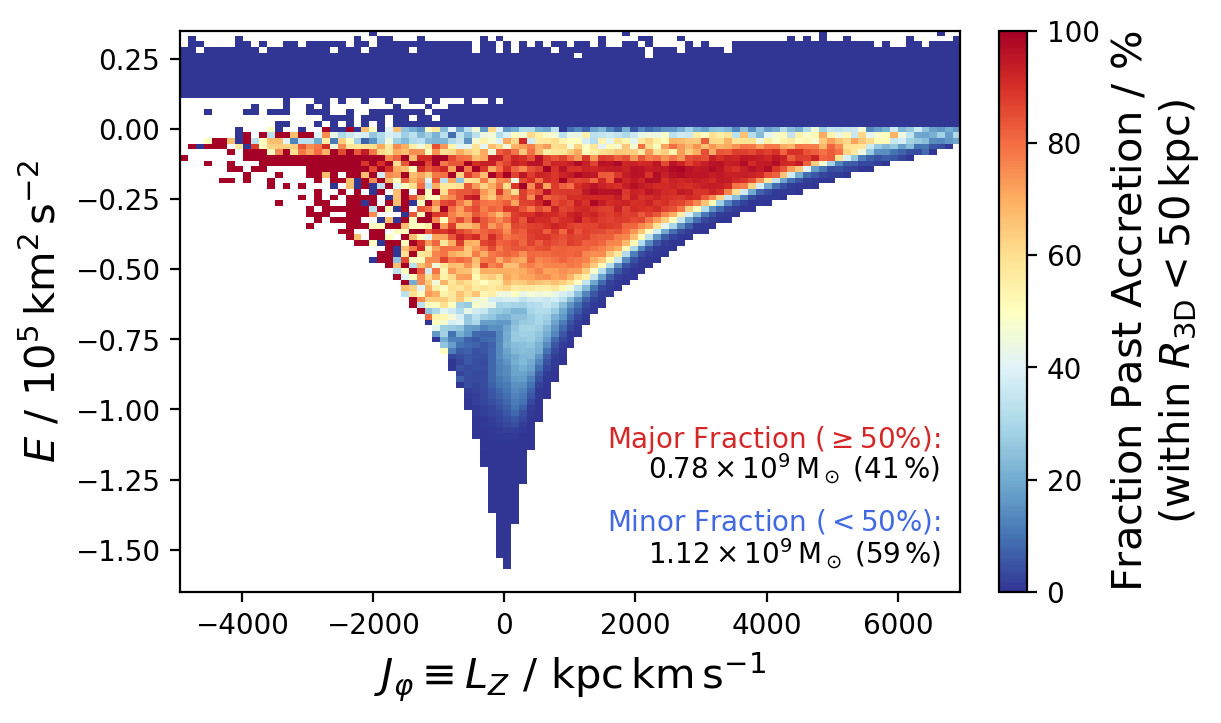}
    \includegraphics[width=0.47\linewidth]{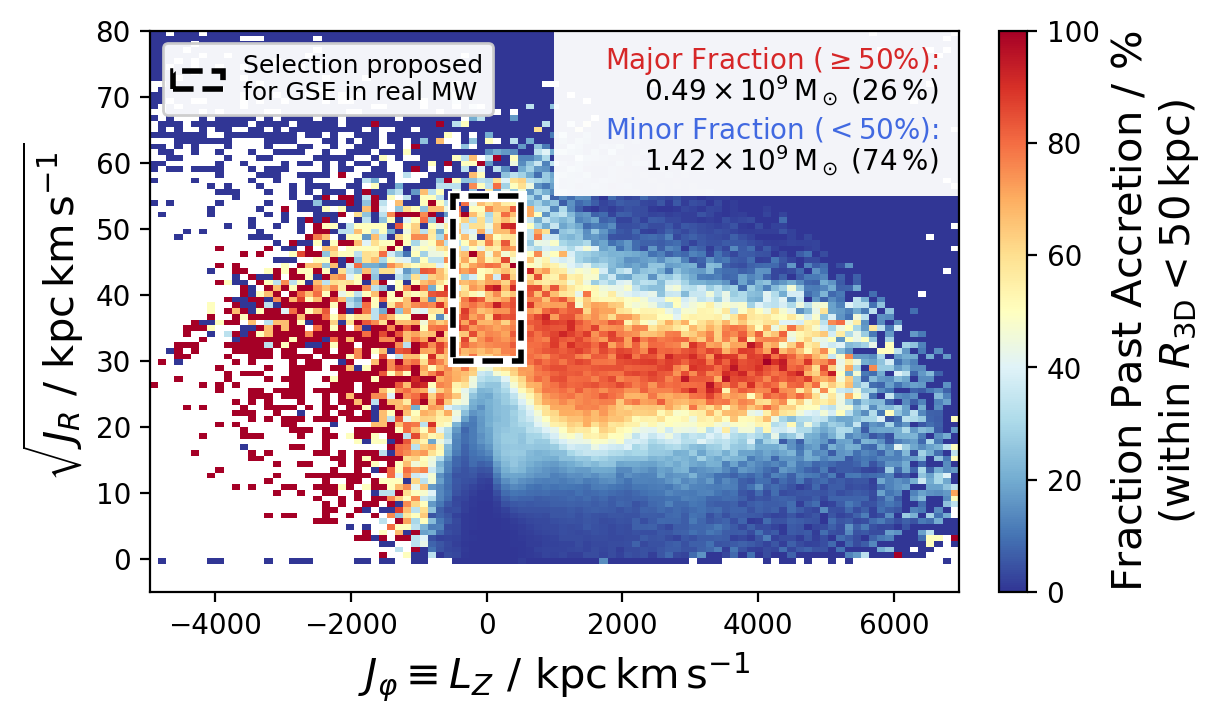}
    \caption{A version of Fig.~\ref{fig:fraction_accreted_in_situ_lz}, but calculating fractions with respect to all star particles \href{https://github.com/svenbuder/golden_thread_I/tree/main/figures}{\faGithub}.}
    \label{fig:fraction_accreted_in_situ_lz_total}
\end{figure*}

\begin{figure*}
    \centering
    \includegraphics[width=0.95\textwidth]{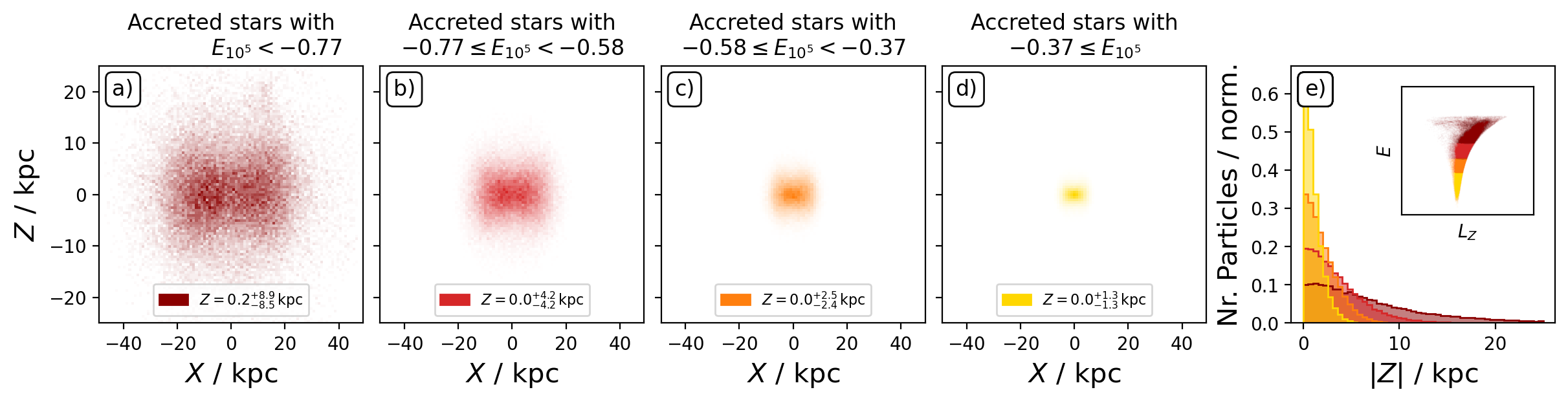}
    \caption{Same as Fig.~\ref{fig:xy_distribution_ezones}, but for $X$-$Z$ dimensions \href{https://github.com/svenbuder/golden_thread_I/tree/main/figures}{\faGithub}.}
    \label{fig:xz_distribution_ezones}
\end{figure*}

\begin{figure*}
    \centering
    \includegraphics[width=0.6\textwidth]{figures/xfe_feh_legend.png}\\
    \includegraphics[width=0.33\textwidth]{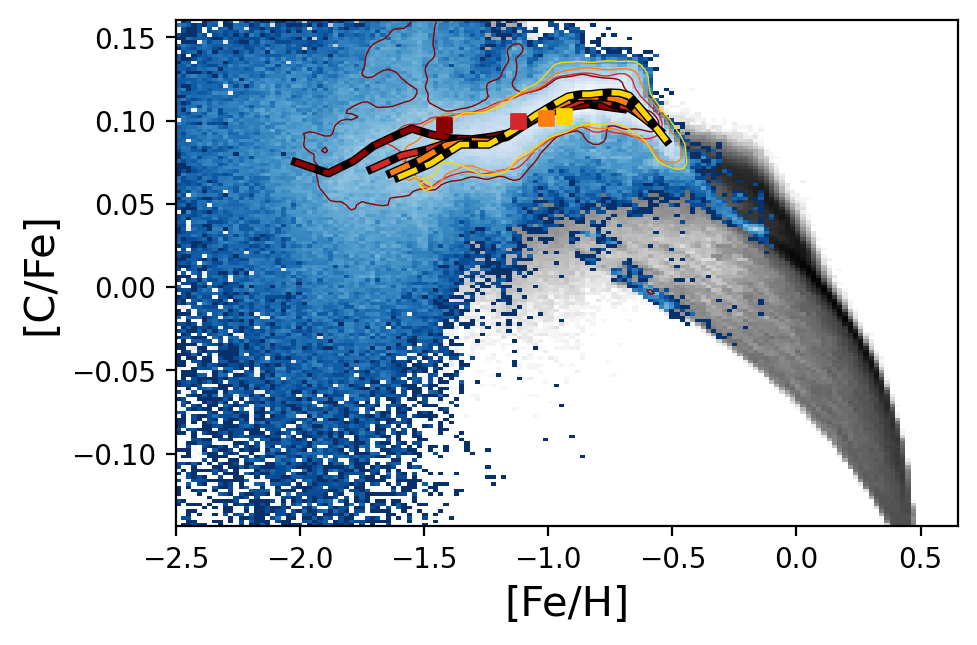}
    \includegraphics[width=0.33\textwidth]{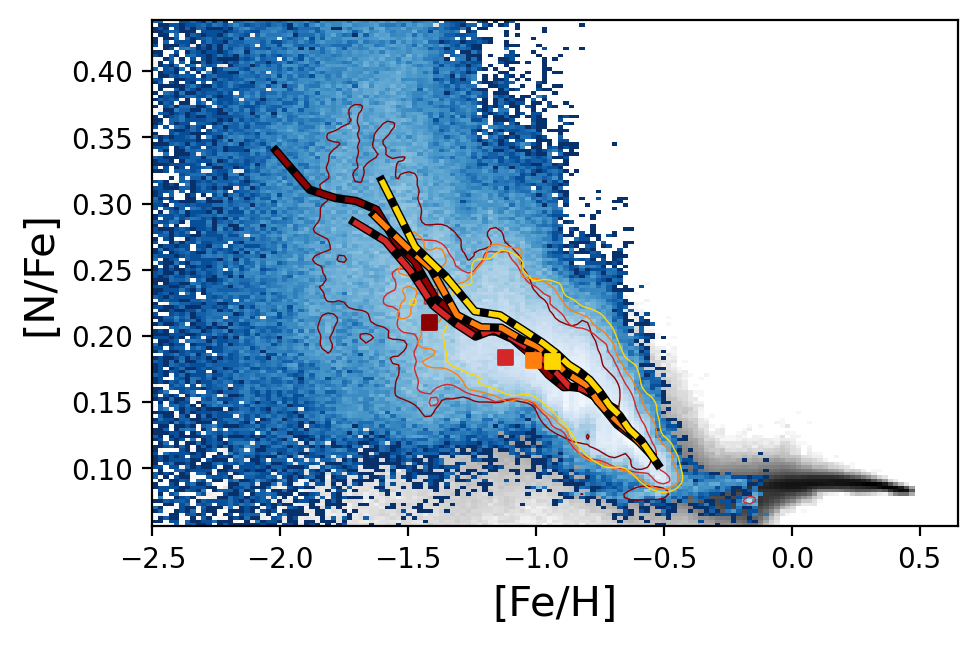}
    \includegraphics[width=0.33\textwidth]{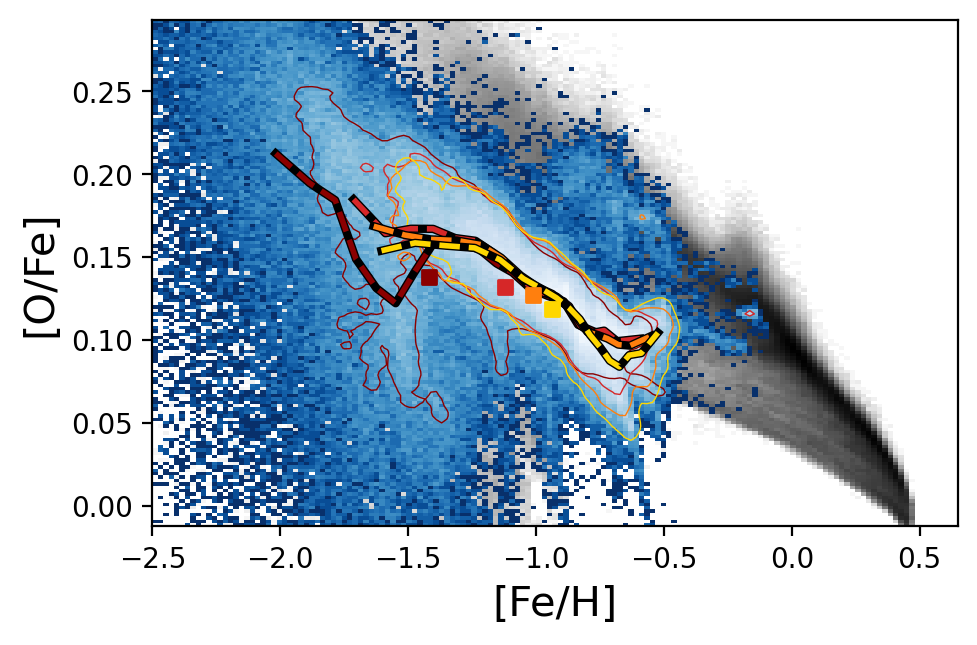}
    \includegraphics[width=0.33\textwidth]{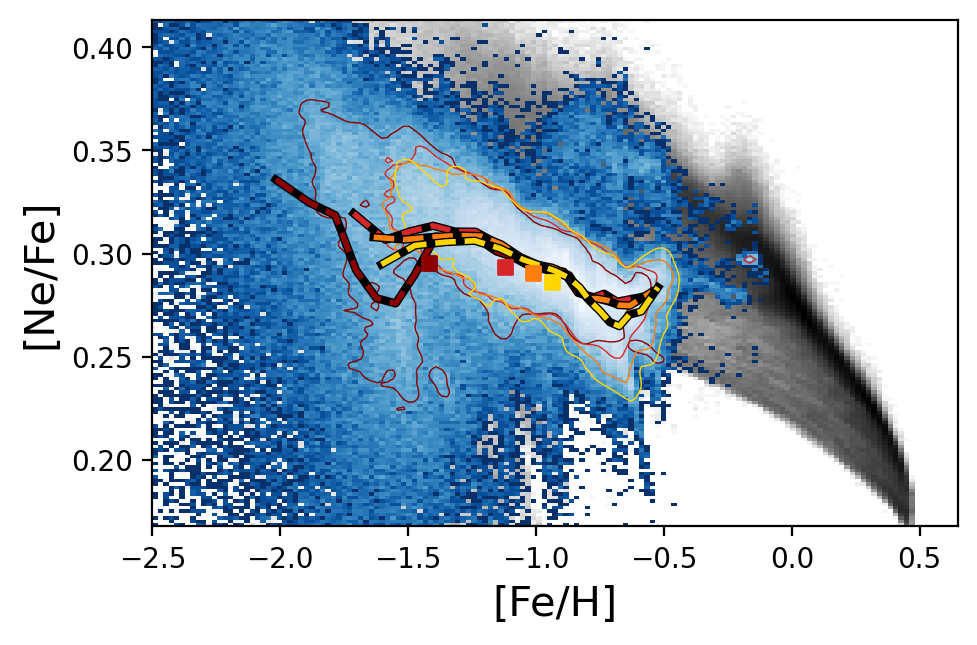}
    \includegraphics[width=0.33\textwidth]{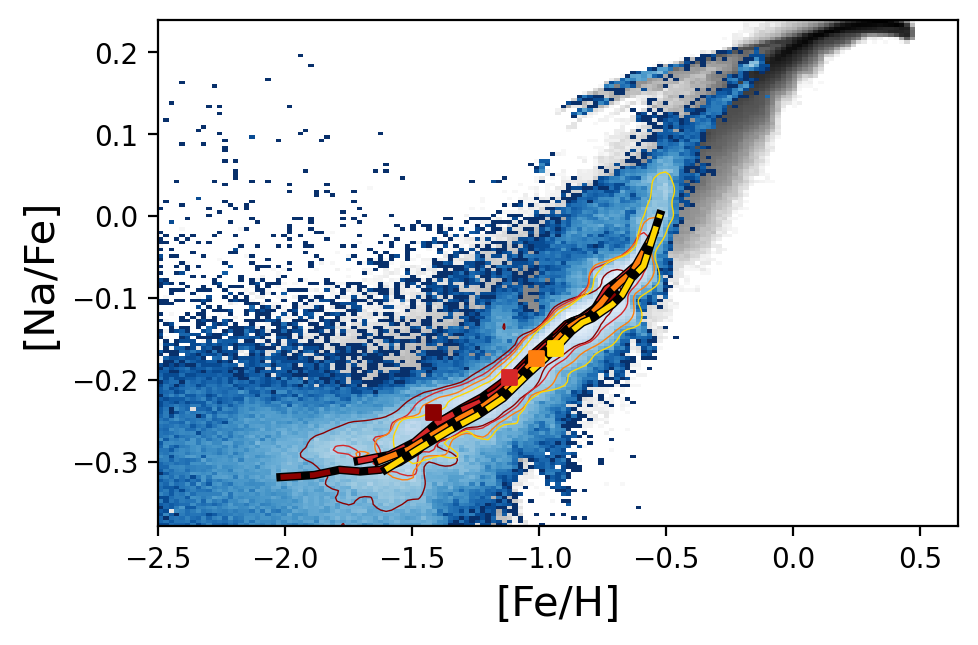}
    \includegraphics[width=0.33\textwidth]{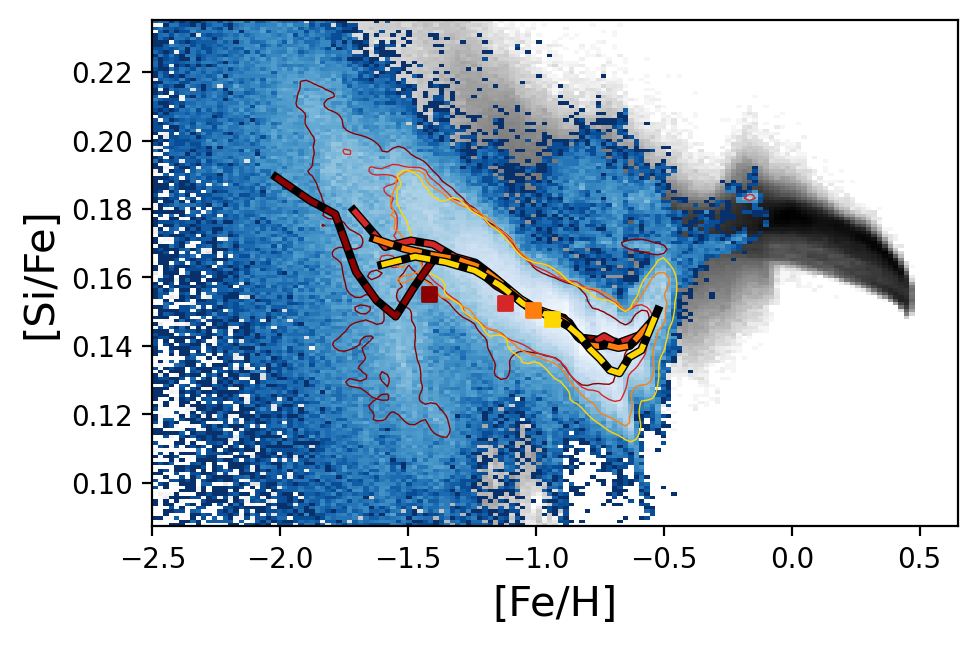}
    \includegraphics[width=0.33\textwidth]{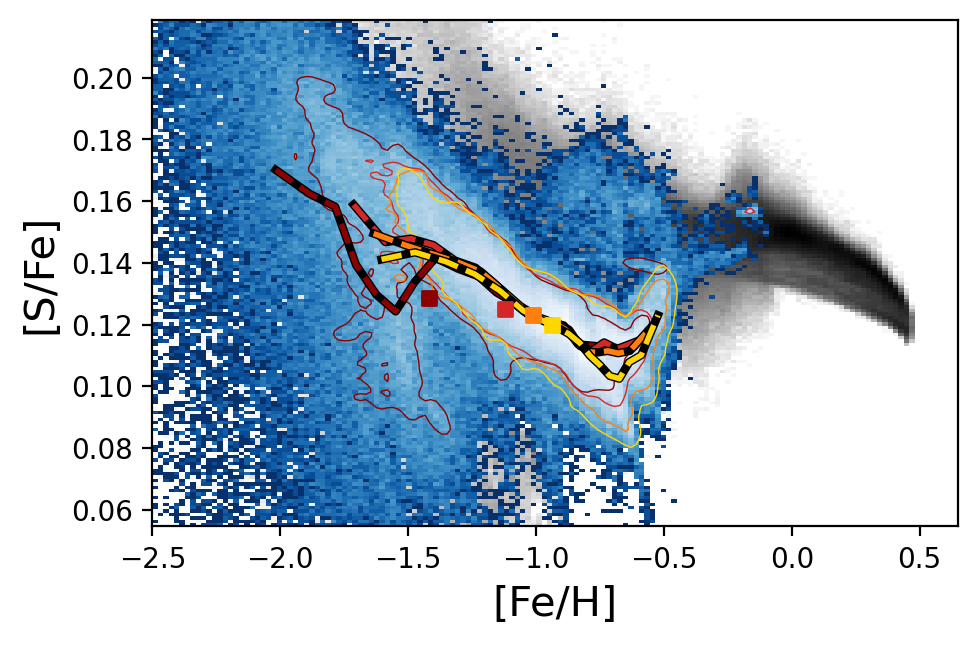}
    \includegraphics[width=0.33\textwidth]{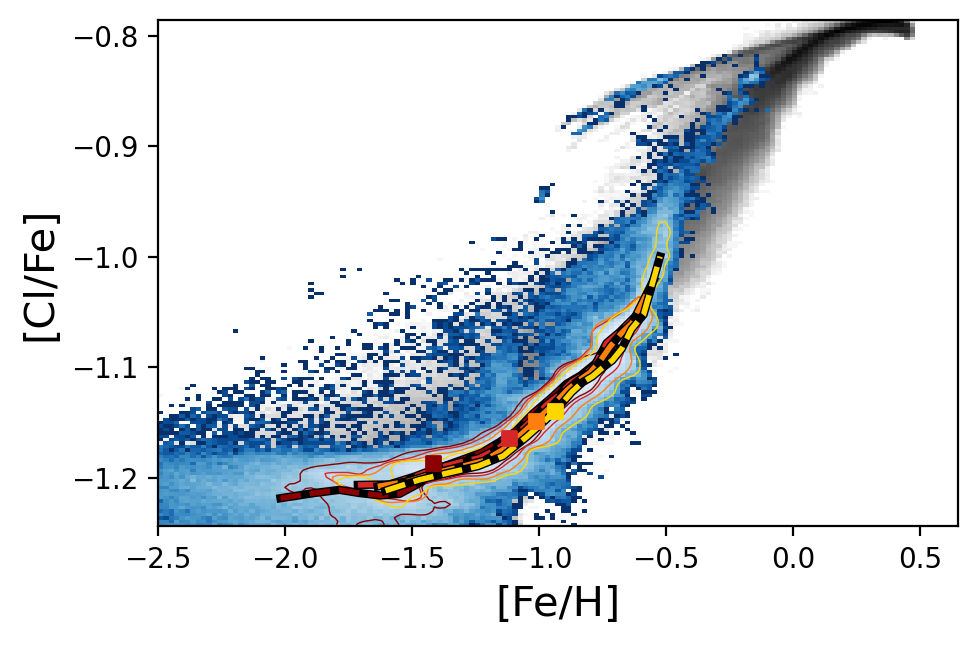}
    \includegraphics[width=0.33\textwidth]{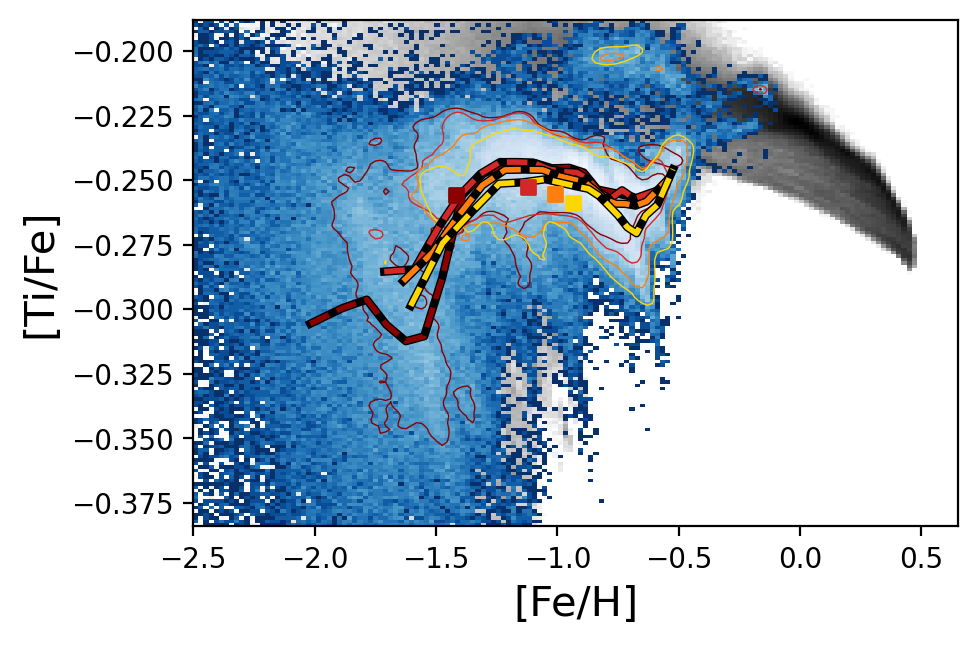}
    \includegraphics[width=0.33\textwidth]{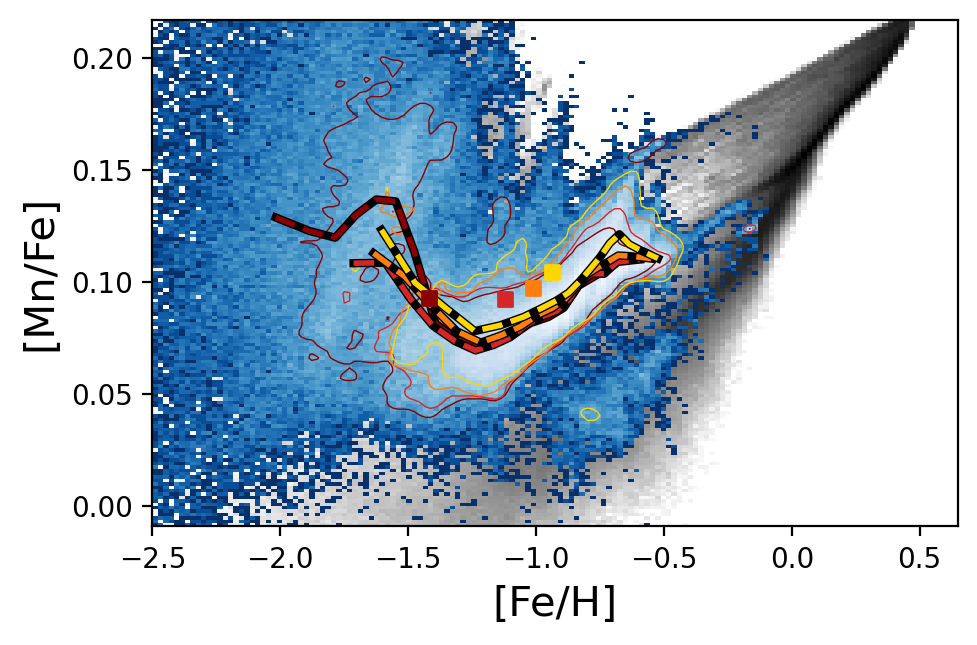}
    \includegraphics[width=0.33\textwidth]{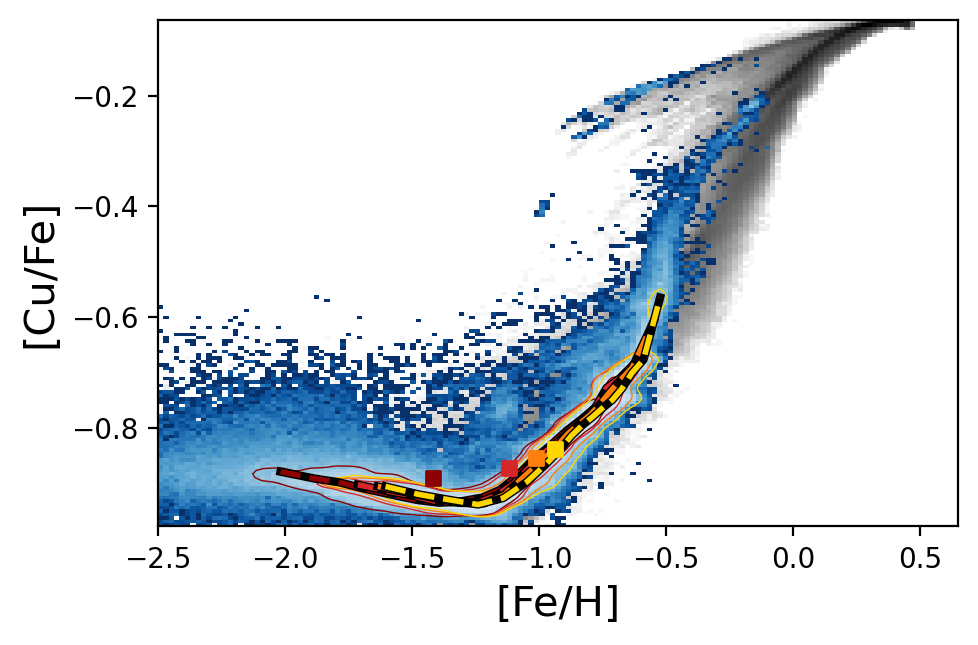}
    \includegraphics[width=0.33\textwidth]{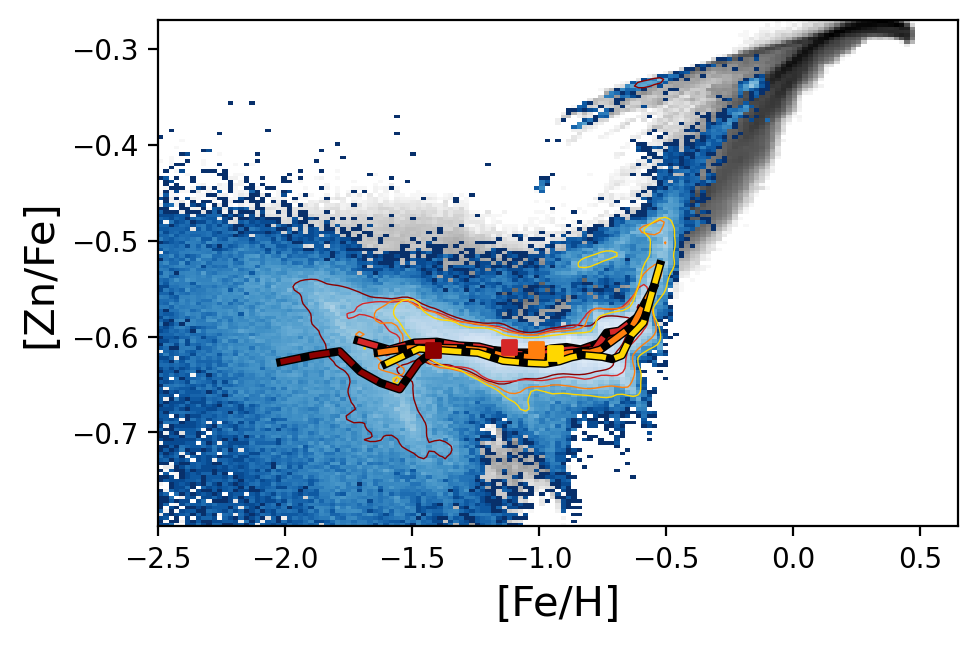}
    \includegraphics[width=0.33\textwidth]{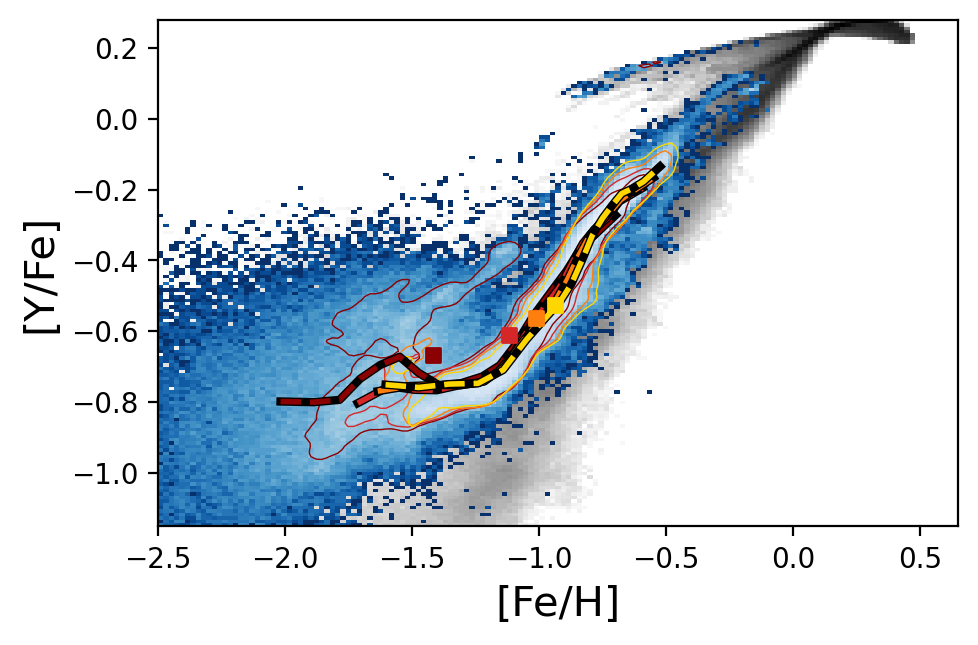}
    \includegraphics[width=0.33\textwidth]{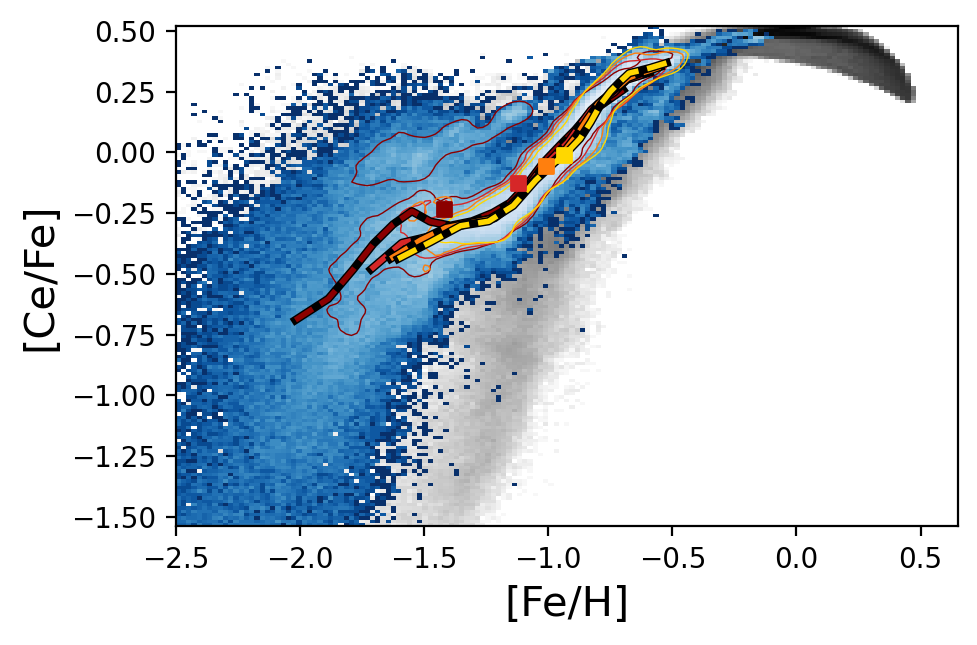}
    \includegraphics[width=0.33\textwidth]{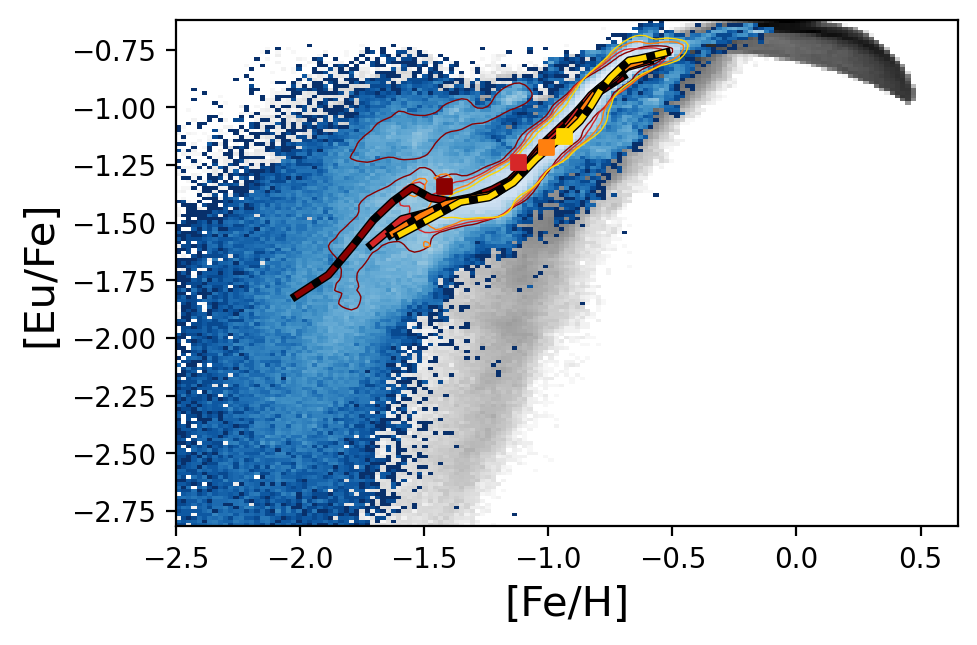}
    \caption{Same as Fig.~\ref{fig:xfe_feh_zones}, for C, N, O, Ne, Na, Si, S, Cl, Ti, Mn, Cu, Zn, Y, Ce, and Eu in the others \href{https://github.com/svenbuder/golden_thread_I/tree/main/figures}{\faGithub}.}
    \label{fig:additional_xfe_feh_zones}
\end{figure*}

\begin{enumerate}[leftmargin=2em,labelwidth=0em]
    \item We provide both the initial selection and resulting average spatial and age ranges of the major accreted structure in Tab.~\ref{tab:birth_position_tabular}.
    \item Fig.~\ref{fig:fraction_accreted_in_situ_lz_total} shows the same distribution of stars in bins of $L_Z$ vs. $E$ and $L_Z$ vs. $\sqrt{J_R}$ as Fig.~\ref{fig:fraction_accreted_in_situ_lz}. In this version, we include ongoing accretion in addition to past accretion and in-situ formation when calculating the fractional contribution for each bin. 
    \item Fig.~\ref{fig:xz_distribution_ezones} shows the spatial distribution of accreted stars as Fig.~\ref{fig:xy_distribution_ezones}, but for the $X-Z$ direction.
    \item Fig.\ref{fig:additional_xfe_feh_zones} includes the chemical distribution of accreted stars beyond the ones already shown in Fig.~\ref{fig:xfe_feh_zones}. They include the alpha-process elements O, Ne, Si, and S (behaving like the alpha-process element Mg), the odd-Z elements Na, Cl, and Cu (behaving like the odd-Z element Al), the iron-peak element Mn (behaving like the iron-peak element Ni), the neutron-capture elements Y, Ce, and Eu (behaving like the neutron-capture element Ba). For completeness, we also include C, N, Ti, and Zn. We omit He, as it scales linearly with [Fe/H]. \changed{We note that \citet[][see their Fig.~15]{Khoperskov2023c} found significantly different trends of sub-Solar [C/Fe] and and sub-Solar [N/Fe] increasing towards Solar values at Solar [Fe/H].}
\end{enumerate}

\label{lastpage}
\end{document}